\documentclass[useAMS,usenatbib]{mn2e}

\usepackage{graphicx}
\usepackage{amsmath}
\usepackage{rotating,booktabs}
\usepackage{graphicx}
\usepackage[T1]{fontenc}
\usepackage[utf8]{inputenc}
\usepackage{mathtools}  
\usepackage{amssymb}
\usepackage{tabulary}
\usepackage{booktabs}
\usepackage{subfigure}
\usepackage{minitoc}
\usepackage[T1]{fontenc}
\usepackage{graphicx}
\usepackage{draftcopy}
\usepackage{algorithm}
\usepackage{algpseudocode}
\usepackage{pifont}
\usepackage{relsize}
\usepackage{amsmath}

\usepackage[T1]{fontenc}
\usepackage[utf8]{inputenc}
\usepackage[usenames,dvipsnames,svgnames,table]{xcolor}
\newcommand{\ee}{\mathrm{e}}

\usepackage[usenames,dvipsnames,svgnames,table]{xcolor}
\usepackage{enumerate}

\newcommand{\ATM}[1]{\textcolor{black}{{#1}}}
\newcommand{\ATMB}[1]{\textcolor{black}{{#1}}}
\newcommand{\ATML}[1]{\textcolor{black}{{#1}}}

\newcommand{\Rd}{\textup{\textrm{d}}}
\usepackage{mwe}    
\usepackage{subfig}
\usepackage[export]{adjustbox}

\usepackage[usenames,dvipsnames,svgnames,table]{xcolor}
\newcommand{\OO}{\mathcal{O}}

\title[Position-dependent PSFs and approximation]{Fast algorithms to approximate the position-dependent point spread function responses in radio interferometric wide-field imaging}
\author[M.  Atemkeng et \textit{al.}]
{M. Atemkeng$^{1}$\thanks{E-mail: m.atemkeng@gmail.com }, O. Smirnov$^{2,3}$, C. Tasse$^{2,4}$, G. Foster$^{5}$ and S. Makhathini$^{2}$\\
$^1$Department of Mathematics, Rhodes University, Grahamstown 6139, South Africa\\
$^2$Department of Physics and Electronics, Rhodes University, PO Box 94, Grahamstown, 6140, South Africa\\
$^3$South African Radio Astronomy Oberservatory, Black River Park, 2 Fir Street, Observatory, Cape Town, 7925, South Africa\\
$^4$GEPI, Observatoire de Paris, CNRS, Universit\'e Paris Diderot, 5 place Jules Janssen, 92190 Meudon, France\\
$^5$University of Oxford, Sub-Department of Astrophysics, Denys Wilkinson Building, Keble Road, Oxford, OX1 3RH, UK}

\begin{document}
\date{Accepted 2020 September 14. Received 2020 September 14; in original form 2020 April 13.}

\pagerange{\pageref{firstpage}--\pageref{lastpage}} \pubyear{2017}

\maketitle

\label{firstpage}

\begin{abstract}
The desire for wide-field of view, large fractional bandwidth, high sensitivity, high spectral and temporal resolution has driven radio interferometry to the point of big data revolution where the data is represented in at least three dimensions with an axis for spectral windows, baselines, sources, etc; \ATMB{where each axis has its own set}
of sub-dimensions. The cost associated with storing and handling these data is very large, and therefore several techniques to compress interferometric data and/or speed up processing have been investigated. 
Unfortunately, averaging-based methods for visibility data compression are detrimental to the data fidelity, since the point spread function (PSF) is position-dependent, i.e. distorted and attenuated as a function of distance from the phase centre. The position dependence of the PSF becomes more severe, requiring more PSF computations for wide-field imaging.
Deconvolution algorithms must take the distortion into account in the major and minor cycles
to properly subtract the PSF and recover the fidelity of the image. 
This approach is expensive in computation   since at each deconvolution iteration a distorted PSF must be computed. We present two algorithms that approximate these position-dependent PSFs with fewer computations. The first algorithm approximates the position-dependent PSFs in the $uv$-plane and the second algorithm approximates
the position-dependent PSFs in the image-plane. The
proposed algorithms are validated using simulated data from the MeerKAT telescope. 
\end{abstract}

\newcommand{\Da}[1]{\textcolor{blue}{{\bf David: #1}}}

\newcommand{\Gcal}{\bmath{\mathcal{G}}}
\newcommand{\GI}{\bmath{\mathcal{GV}}}
\newcommand{\Rcal}{\bmath{\mathcal{R}}}
\newcommand{\Mcal}{\bmath{\mathcal{M}}}
\newcommand{\Fcal}{\bmath{\mathcal{F}}}
\newcommand{\GG}{\mathcal{G}}
\newcommand{\CC}{\mathcal{C}}
\newcommand{\KKK}{\mathcal{K}}
\newcommand{\BB}{B}
\newcommand{\EEE}{\bmath{E}}
\newcommand{\GGG}{\bmath{G}}
\newcommand{\Vector}{\bmath{u}}
\newcommand{\VV}{\mathcal{V}}
\newcommand{\VVVVV}{V}
\newcommand{\VVMT}{\bmath{V}}
\newcommand{\VVSC}{V}
\newcommand{\WW}{\mathcal{W}}
\newcommand{\II}{\mathcal{I}}
\newcommand{\IIA}{\widetilde{\mathcal{I}}}
\newcommand{\argmax}{\operatornamewithlimits{argmax}}
\newcommand{\IID}{\mathcal{I}^\mathrm{D}}
\newcommand{\IIDI}{\mathcal{I}^\mathrm{DI}}
\newcommand{\EE}{\mathcal{E}}
\newcommand{\FF}{\mathcal{F}}
\newcommand{\HH}{\mathcal{H}}
\newcommand{\TT}{\mathcal{T}}
\newcommand{\NN}{\mathcal{N}}
\newcommand{\uu}{\bmath{u}}
\newcommand{\Btf}{\mathsf{B}^{[\Delta t\Delta\nu]}}
\newcommand{\Btfleft}{\mathsf{B}^{\mathrm{lft}}}
\newcommand{\Btfright}{\mathsf{B}^{\mathrm{rgt}}}
\newcommand{\Babtf}{\mathsf{B}^{[\alpha\Delta t,\beta\Delta\nu]}}
\newcommand{\BOVERLAP}{\mathsf{B}^{[\Delta^\mathrm{olp} t,\Delta^\mathrm{olp}\nu]}}
\newcommand{\Bab}{\mathsf{B}^{[\alpha\beta]}}
\newcommand{\Buv}{\mathsf{B}^{[uv]}}
\newcommand{\Bij}{\mathsf{B}}
\newcommand{\Ptf}{\Pi^{[t\nu]}}
\newcommand{\Puv}{\Pi^{[uv]}}
\newcommand{\rrm}{\widetilde{\eta}}

\newcommand{\Vapp}{V^\mathrm{(dis)}}
\newcommand{\Vmf}{\mathcal{V}^\mathrm{(m)}}
\newcommand{\Vs}{V^\mathrm{(s)}}
\newcommand\norm[1]{\left\lVert#1\right\rVert}
\newcommand{\Bda}{\mathsf{B}^{[\Delta_{pq} t, \Delta_{pq} \nu]}}
\newcommand{\Bdashortest}{\mathsf{B}^{[\Delta_{34} t]}}
\newcommand{\PP}{\mathcal{P}}

\newcommand{\Bdaph}{\mathsf{B}^{[\Delta_{\alpha \beta} t, \Delta_{\alpha \beta} \nu]}}
\newcommand{\Bdaphfreq}{\mathsf{B}^{[\Delta_{pq} t, \Delta_{pq}\nu]}}
\newcommand{\Dbda}{\mathsf{D}^{[\Delta_{pq} t, \Delta_{pq} \nu]}}
\newcommand{\Dbdatime}{\mathsf{D}^{[\Delta_{pq} t]}}
\newcommand{\Dbdafreq}{\mathsf{D}^{[\Delta_{pq} \nu]}}
\newcommand{\Dbdaph}{\mathsf{D}^{[\Delta_{\alpha \beta} t, \Delta_{\alpha \beta} \nu]}}
\newcommand{\Dbdaphtime}{\mathsf{D}^{[\Delta_{\alpha \beta} t]}}
\newcommand{\Dbdaphfreq}{\mathsf{D}^{[ \Delta_{\alpha \beta} \nu]}}
\newcommand{\Dbdlong}{\mathsf{D}^{[\Delta_{12} t, \Delta_{12} \nu]}}
\newcommand{\Dbdmedium}{\mathsf{D}^{[\Delta_{23} t, \Delta_{23} \nu]}}
\newcommand{\Dbdshort}{\mathsf{D}^{[\Delta_{34} t, \Delta_{34} \nu]}}
\newcommand{\Bdaphlong}{\mathsf{B}^{[\Delta_{12} t, \Delta_{12} \nu]}}
\newcommand{\Bdaphmedium}{\mathsf{B}^{[\Delta_{23} t, \Delta_{23} \nu]}}
\newcommand{\Bdaphshort}{\mathsf{B}^{[\Delta_{34} t, \Delta_{34} \nu]}}
\newcommand{\VVM}{\mathcal{V}^\mathrm{M}}
\newcommand{\RR}{\mathcal{R}}
\newcommand{\XX}{\mathcal{X}}
\newcommand{\VVME}{\mathcal{V}^\mathrm{ME}}
\newcommand{\VVDG}{\mathcal{V}^\mathrm{degrid}}
\newcommand{\VVG}{\mathcal{V}^\mathrm{grid}}
\renewcommand{\theenumi}{\Alph{enumi}}
\newcommand{\DDR}{d}
\newcommand{\Wm}{\boldsymbol{W}}
\newcommand{\Vtm}{\boldsymbol{V}}
\newcommand{\Va}{V}
\newcommand{\EDIT}[1]{\textcolor{black}{{#1}}}
\newcommand{\PSF}{\mathcal{P}}
\newcommand{\DIT}{\mathcal{K}}
\newcommand{\QQ}{\mathcal{P}^{(dis)}}
\newcommand{\Cm}{\boldsymbol{C}}
\newcommand{\Vmm}{\boldsymbol{V}_{\mu}^{\mathrm{M}}}
\newcommand{\FFF}{\boldsymbol{F}}
\newcommand{\Wa}{W}
\newcommand{\FFFH}{\boldsymbol{F}^{H}}
\newcommand{\AAA}{\boldsymbol{A}}
\newcommand{\IIDT}{\mathcal{I}^\mathrm{}}
\section{Introduction to the broad problem}
\label{introduction}
New radio interferometric arrays produce large volumes of data that has to be transported over large distances for processing. The MeerKAT \citep{jonas2009meerkat} and \ATML{the} LOFAR \citep{van2013lofar} telescopes are examples of the current state of the art. With the 64 antenna stations of 
the MeerKAT telescope located at the Karoo desert in South Africa the data volume to be transmitted to the correlation station is around anywhere from 64 Gb/s to 0.5 Pb/s. 
The \ATM{Square Kilometre Array} (SKA, \citet{dewdney2009square})  is a future wide-field of view, large fractional bandwidth, high sensitivity, high spectral and temporal resolution imaging instrument designed to image the sky at \ATML{arcsecond} angular resolution even at low frequencies~\citep{labate2017ska1}. 
The SKA is expected to produce a data flow of the order of Pb/s and the data will be transmitted between the SKA \ATM{partner countries} i.e. over distances at the scale of the \lq\lq Earth radius’’. 
Experience with these currently operational
big data radio interferometer \ATM{arrays} shows an increase in computational complexity for transmitting, storing and processing the data. 
At the SKA scale, even using the most powerful supercomputers,  the computation will still remain a significant challenge.
During interferometric data acquisition the signal is corrupted by various effects including turbulence from the atmosphere, noise from the instrument and  sparse sampling of the Fourier coefficients  of the measured sky distribution.  
 The processing steps include calibrating the raw \ATM{visibility data} to remove these corruptions, and imaging the data while mitigating the impact of the sparse Fourier coefficient sampling in the $uv$-plane.
The step that mitigates or compensates for  the unsampled Fourier coefficients in the $uv$-plane is known as image deconvolution and  results in improving the quality and the signal to noise ratio (S/N) of the image. 

\subsection{Deconvolution and decorrelation}
Deconvolution is a well-known image reconstruction operation in radio interferometry. Its classical variants include CLEAN and maximum entropy-based algorithms~\citep{hogbom1974aperture,bhatnagar2004scale,offringa2014wsclean,ables1974maximum}.
New reconstruction methods like compressive sensing \citep{carrillo2014purify, dabbech2015moresane} and Bayesian inference~\citep{junklewitz2016resolve}  are promising techniques when compared to results from the  CLEAN algorithms. All these algorithms, however, aim
to predict or compensate for the unsampled regions in the observed \ATM{visibility data}, and this prediction is becoming increasingly challenging in processing with the big data nature of these new instruments. To speed up the processing in this era of big data, new data size reduction strategies for radio interferometric data compression must be developed. 
 This motivates recent work on: 
 \begin{itemize}
  \item Baseline-dependent window functions~\citep{atemkeng2016} and baseline-dependent averaging~\citep{wijnholds2018baseline, atemkeng2018baseline} to compress radio interferometric data while minimising the loss of sources amplitude across the field of view.
  \item Lossy compression \ATM{for  radio interferometric  data}~\citep{offringa2016compression}, and  online imaging strategies~\citep{cai2017online} where the observed \ATM{visibility data} are imaged row by row as they are acquired which does not require saving the entire data of the observation.
  \item \ATML{Visibility} distribution~\citep{meillier2018distribution}; the big data is split into small blocks of data and shared over several processing nodes where the image reconstruction is performed in parallel.
  \item The Fourier dimensional reduction technique~\citep{vijay2017fourier} where methods like the singular value decomposition~\citep{golub1970singular} and random projection~\citep{bingham2001random} are applied to compress the gridded data during imaging and deconvolution. 
 \end{itemize}
All these new compression algorithms are detrimental to the image quality and fidelity which, if traditional visibility averaging or baseline-dependent averaging
are used, cause the visibilities to decorrelate. This results in changes of the local PSF for each source in the image~\citep{atemkeng2016data,tasse2018faceting, bonnassieux2020decoherence}.
Each of these local PSFs is attenuated in amplitude and smeared in shape differently.

Deconvolution with the classical CLEAN algorithms is an iterative approach; each iteration consists of finding the brightest pixel value in the image which is then convolved \ATMB{with} the \ATM{effective PSF (\ATM{i.e. PSF at the phase centre of the observation})} before the result of the convolution is subtracted from the image. 
Using the effective PSF  in place of these local PSFs to deconvolve all sources in the image introduces smearing artifacts around these sources which can significantly decrease the overall image S/N and bias the morphology of sources in the deconvolved image.
An acceptable solution when considering the effective PSF to deconvolve all these sources in the image is to consider correlating or averaging the data with a very small channel width and integration time in such a way that the local PSF of each of the sources in the image is no longer distorted.
\ATM{It is impractical to \ATMB{keep} the time and frequency resolution of the visibilities sufficiently high  as this comes with  massive data volumes and therefore high computational demands in post-processing.}
To avoid the massive volume of data the channel width and integration time need to be fixed accordingly but the resulting local PSF of each source in the image will be attenuated and smeared. Also, the data can be further simple-averaged or baseline-dependent averaged still to reduce the data volume and speed up processing which will further distort these local PSFs differently. 

To account for the distortion or to correct for the decorrelation \ATMB{introduced} by averaging the visibilities,  an ideal solution would be to deconvolve each source in the image with its own local PSF. This implies that at each deconvolution step a local PSF must be evaluated.
 A solution similar to this approach is implemented in DDFacet~\citep{tasse2018faceting}; a CLEAN and faceting based deconvolution imaging framework. 
 The local PSF of each source within a small facet does not vary that much, thus, the negative effect of decorrelation can be taken as negligible. \ATM{As such, the DDFacet imager internally evaluates the local PSF at the centre of each small facet by brute-force and uses this local PSF to deconvolve all 
 sources within the facet. The brute-force computation of the local PSFs per facet is not expensive as  opposed to per source. 
What if the facets are large?} The local PSF of each source within a large facet will vary significantly \ATM{which then requires} that one should compute 
all these local PSFs within the facet: it is costly in term of computing.
For a non-faceting imager and for wide-field imaging, evaluating these local PSFs by brute-force is a very complex task to handle given that the computational requirements would increase linearly with the size of the image to deconvolve and the number of sources in the image. To remove this computational restriction we propose to approximate these local PSFs with fewer computation compared to a brute-force approach. The discussion in this paper is thus limited to
how to approximate these local PSFs, rather than the actual deconvolution algorithm itself. 

\ATM{Another category of imaging artifacts are distant source sidelobes. These artifacts are generated by the sidelobes of bright objects
extending inside the interferometer array \ATMB{field of view}. These objects are sometimes found within the interferometer array field of view. A common example is the bright radio galaxy
Cygnus A~\citep{boccardi2016stratified}. To overcome this problem with LOFAR observations, a catalogue of bright sources (A-team sources) that can contaminate observations has
been established. The A-team sources are often subtracted from observations before analysis. One common technique for source subtraction used in the literature is peeling \citep{smirnov2011revisiting}, which solves for direction-dependent gains across the \ATMB{field of view}. An accurate subtraction of these bright sources can significantly improve deconvolution in wide-field imaging. \ATMB{Once these bright sources (or the sidelobes if the bright sources are out of the field of view)} are properly subtracted from the observing field, the dynamic range required to make an image of the sources of interest in the observing field is reduced, which eventually reduces the requirements on the accuracy of the PSF used for deconvolution.  
Note that the observed and distorted point source is the unnormalised local PSF that we want to approximate and that can be used in the $uv$-plane \ATML{to properly} subtract the point sources. The subtraction in the $uv$-plane does not require that the source is deconvolved to extract the predicted model before the  subtraction is carried out.} Throughout this work, 
we refer to these distorted local PSFs as position-dependent PSFs.

\subsection{Contribution and manuscript organisation}
This work focuses on the approximation of the position-dependent PSFs across the image which requires fewer computing resources.  The proposed methods use the \ATM{visibility measured at the centre of the averaging interval} and
the effective PSF to establish two mathematical frameworks that approximate the position-dependent PSFs across the image. Compared with the brute-force computing approach of the position-dependent PSFs which uses the entire observed data to compute each \ATM{position-dependent} PSF, our two algorithms work independently as follows:
\begin{itemize}
 \item In the $uv$-plane, the first algorithm  \ATMB{uses the phase of the visibility acquired at the centre of each averaging interval as \ATML{an approximation} of the phase of the averaged visibility (i.e. the phase gradient). Throughout
this work, the phase gradient refers to the phase of the visibility acquired at the centre of
the averaging interval.}
  \item In the image-plane, the second algorithm computes the effective PSF once, approximates some decorrelation coefficients of each source and \ATM{applies} this to the effective PSF.
\end{itemize}
The first algorithm is limited to the number of phase gradients rather than the entire observed \ATM{visibility data} and the second algorithm computes the well-known effective PSF once and \ATM{uses this to approximate} the position-dependent PSFs rather than computing all these position-dependent PSFs individually by brute-force. These two algorithms thus show an increase in computing efficiency, and with the significant confirmation that the error introduced by the approximation is negligible compared to the brute-force approach. 

The rest of this work is organised as follows: in Section~\ref{sect2},  a mathematical model to understand the position-dependent PSFs is proposed. The mathematical formulations are well documented in radio interferometry \ATM{literature~\citep{thompson1999fundamentals}} but it is useful to present them for subsequent use; we start the formulation from the visibilities of the entire sky and then restrict this to the visibilities of a 
single point source. 
Section~\ref{fatsderivationspeudopsfs} proposes the two algorithms for approximating the position-dependent PSFs and their computational complexity \ATM{are discussed in detail}. The algorithms are tested and compared to the brute-force method in Section~\ref{simulations} using simulated data from the MeerKAT telescope, and Section~\ref{conclusion} concludes the  work. 

\section{Imaging and position-dependent PSFs}
\label{sect2}
This section introduces the notion of imaging and effective PSF that are relevant to this work. The section also discusses the mathematical frameworks that describe the position-dependent PSFs.
\subsection{Imaging}
\newcommand{\IIGRAL}{\mathop{\mathlarger{\iint}}}
\newcommand{\IGRAL}{\mathop{\mathlarger{\int}}}
\newcommand{\LL}{\mathbf{l}}
Following the van Cittert-Zernike theorem~ \citep{thompson1999fundamentals,thompson2001fundamentals}, and assuming  no sampling and other corruption effects, the visibility function  under specific conditions (see \citet{thompson2001fundamentals}) is given by the 2D Fourier relationship:
\begin{alignat}{2}
\mathcal{V}&=\int \int \II \ee^{-2 i\pi\bmath{u}_{} \bmath{l}} \Rd \bmath{l}\\
	   &= \mathcal{F}\ATM{\{}\II\ATM{\}},
			  \label{4.2.1}
\end{alignat}
where $\bmath{l}=(l,m)$ is the sky position with $l$ and $m$  the direction cosines.  The components $u$ and $v$ of the vector 
$\bmath{u}_{}=(u, v)$ describe the separation between two antenna elements \ATM{referred to as} a baseline with $u$ \ATMB{aligned with} east-west and $v$ \ATMB{with} south-north.  \ATM{The baseline is measured in wavelengths and can be treated as a function of frequency $\nu$ and time $t$: $\bmath{u}_{}=\bmath{u}_{}(t,\nu)=\bmath{u}_{}(t)\nu/c$ where $\bmath{u}_{}(t)$ is in metre and $c$ the speed of light.}
Here, $\II$ is the \ATM{apparent} sky and $\mathcal{F}^{}\ATM{\{\cdot\}}$ represents a 2D Fourier transform operator and $\mathcal{F}^{-1}\ATM{\{\cdot\}}$ will represent its inverse throughout this work.

 The visibility that an interferometer array measures \lq\lq the measured visibility’’ is the 
sampled version  of $\mathcal{V}$  at each baseline, and discrete time-frequency bin.  Unfortunately, due to the discrete sampling inverting the measured visibility results \ATMB{in} the so called \lq\lq dirty image’’, $ \IID$, and not the \ATM{apparent} image of the sky:
\begin{alignat}{2}
\IID&=\mathcal{F}^{-1}\ATM{\big\{}\mathcal{W}_{}\mathcal{V}_{}\ATM{\big\}}\label{sect:2x} \\
&= \II\circ\mathcal{P},\label{sect:2}
\end{alignat}
where the symbol $\circ$ denotes the convolution operator and $\mathcal{W}$ is the weighted sampling function in the extent of the $uv$-plane.  We note that $\mathcal{P}$ is the inverse Fourier transform of the weighted sampling function, i.e.  $\mathcal{P}=\mathcal{F}^{-1}\ATM{\{}\mathcal{W}\ATM{\}}$  which is the resulting PSF of the observation.
Eq.~\ref{sect:2} is the familiar result which shows that the dirty image is the result of the \ATM{apparent} sky \ATMB{convolved with} the PSF of the observation.

\subsection{Effective PSF}
\label{sect2.2}
Making use of the convolution definition,  Eq.~\ref{sect:2}  can be  \ATMB{expressed as a direction-dependent convolution:}
\begin{alignat}{2}
\IID &= \int \int \II(\bmath{l}_0)\mathcal{P}( \bmath{l}_0, \bmath{l}-\bmath{l}_0)\Rd \bmath{l}_0.\label{sect:4.2.8}
\end{alignat}
Current deconvolution algorithms based on CLEAN use
the effective PSF $\mathcal{P}( \bmath{0}, \bmath{l})$, i.e. the PSF at the phase centre $\bmath{l}_0=\bmath{0}=(0,0)$ to deconvolve all sources in the 
 \ATM{dirty image}. Using $\mathcal{P}( \bmath{0}, \bmath{l})$ as an \ATML{approximation} of  \ATMB{$\mathcal{P}( \bmath{l}_0, \bmath{l}-\bmath{l}_0)$ for all $\bmath{l}_0$} holds only if $\bmath{l}_0$ is at the phase centre proximity:
 \begin{alignat}{2}
  \displaystyle{\lim_{\bmath{l}_0 \to \bmath{0}}\mathcal{P}( \bmath{l}_0, \bmath{l}-\bmath{l}_0)}&= \mathcal{P}( \bmath{0}, \bmath{l}).\label{eq:xxx}
 \end{alignat}
Wide-field imaging is explicitly the domain where $\bmath{l}_0$ is assumed to be large, and so this condition no longer \ATML{holds}. It is therefore,
by definition, outside of the regime of validity for this hypothesis.

In reality, during observations the weighted sampling
function $\mathcal{W}$ is a set of \ATMB{weighted  delta} functions $\delta$
where the sampling rate depends on the integration time and
the width of the channels. The extent of the weighted sampling  function  in the entire $uv$-plane depends on the number of antenna elements, the  total length of observing period, the total bandwidth and  the interferometer array layout:
\begin{equation}
\mathcal{W}= \sum_{pqkr}^{}W_{pqkr} \delta_{pqkr}, \label{sum4.2.7}
\end{equation}
where  $W_{pqkr}$ is the weight applied to the sampled visibility for baseline $pq$ at discrete-time and frequency indexed by $k$ and $r$ respectively. 
Here, $\delta_{pqkr}(\bmath{u}_{})=\delta_{}(\bmath{u}_{}-\bmath{u}_{pqkr})$ is the Delta function shifted to the sampled point $pqkr$ with
\begin{alignat}{2}
 \bmath{u}_{pqkr}&=\bmath{u}_{pq}(t_k,\nu_r)
 &=(u_{pqkr}, v_{pqkr}).
\end{alignat}
We can derive $\mathcal{P}( \bmath{0}, \bmath{l})$  \ATMB{by simply taking the inverse Fourier transform of Eq.~\ref{sum4.2.7}}, i.e.
\begin{alignat}{2}
\begin{split}
 \mathcal{F}^{-1}\ATM{\{}\mathcal{W}\ATM{\}}&=\sum_{pqkr}^{}W_{pqkr}\mathcal{P}_{pqkr}\\
				     &= \mathcal{P}( \bmath{0}, \bmath{l}),\label{eq:xxPe}
\end{split}
\end{alignat}
where  $\mathcal{P}_{pqkr}=\mathcal{F}^{-1}\ATM{\{}\delta_{pqkr}\ATM{\}}$. We note that $W_{pqkr}\mathcal{P}_{pqkr}$ is the effective PSF representing the inverse Fourier transform of \ATMB{the weighted visibility} sample at $pqkl$.
Eq.~\ref{eq:xxPe}  does \textbf{not}  result \ATMB{in} a position-dependent 
PSF  but rather the \ATMB{effective} PSF.

\begin{figure*}
\centering
       \includegraphics[width=.35\linewidth]{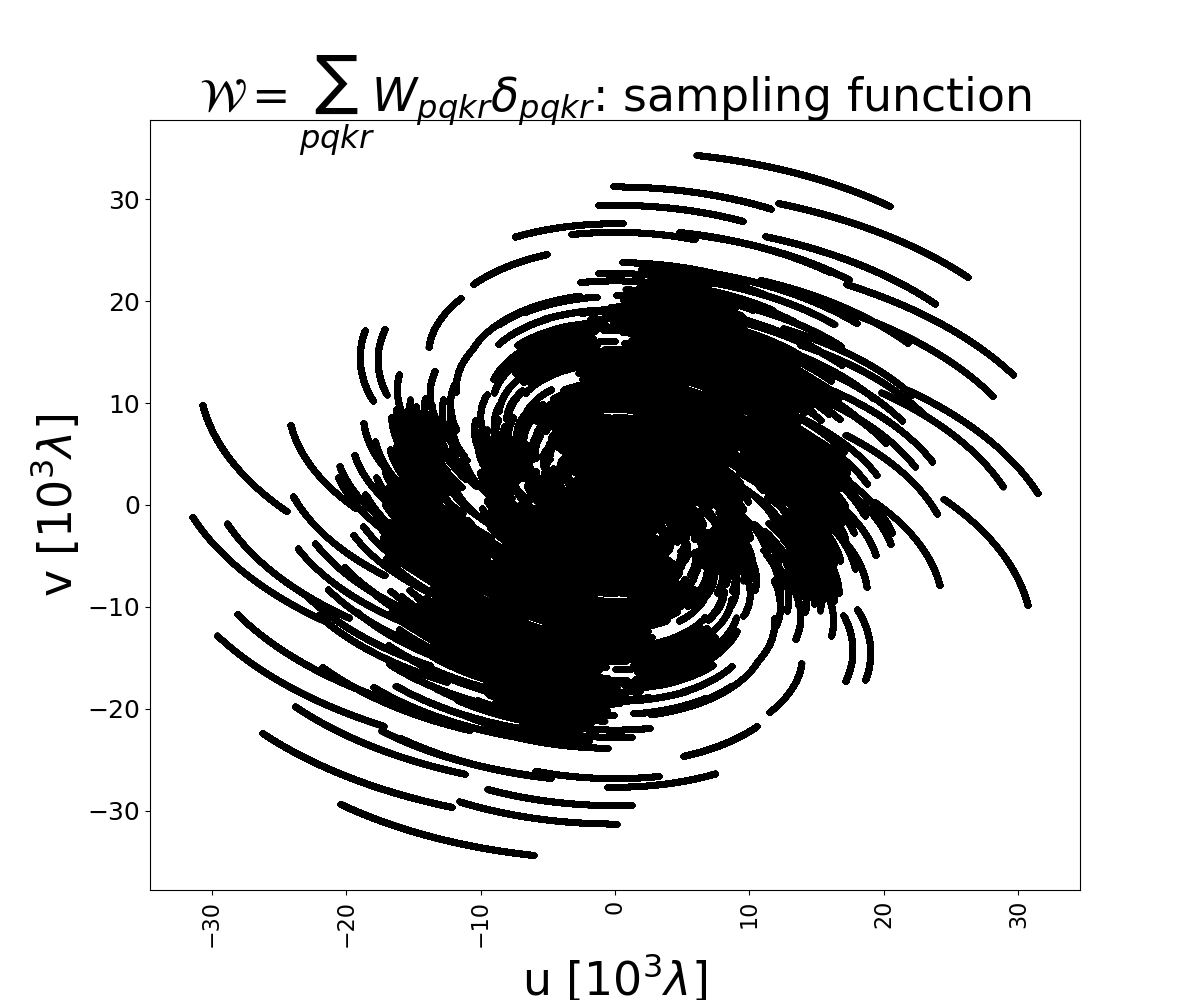}\hspace{-0.7cm}
          \includegraphics[width=.35\linewidth]{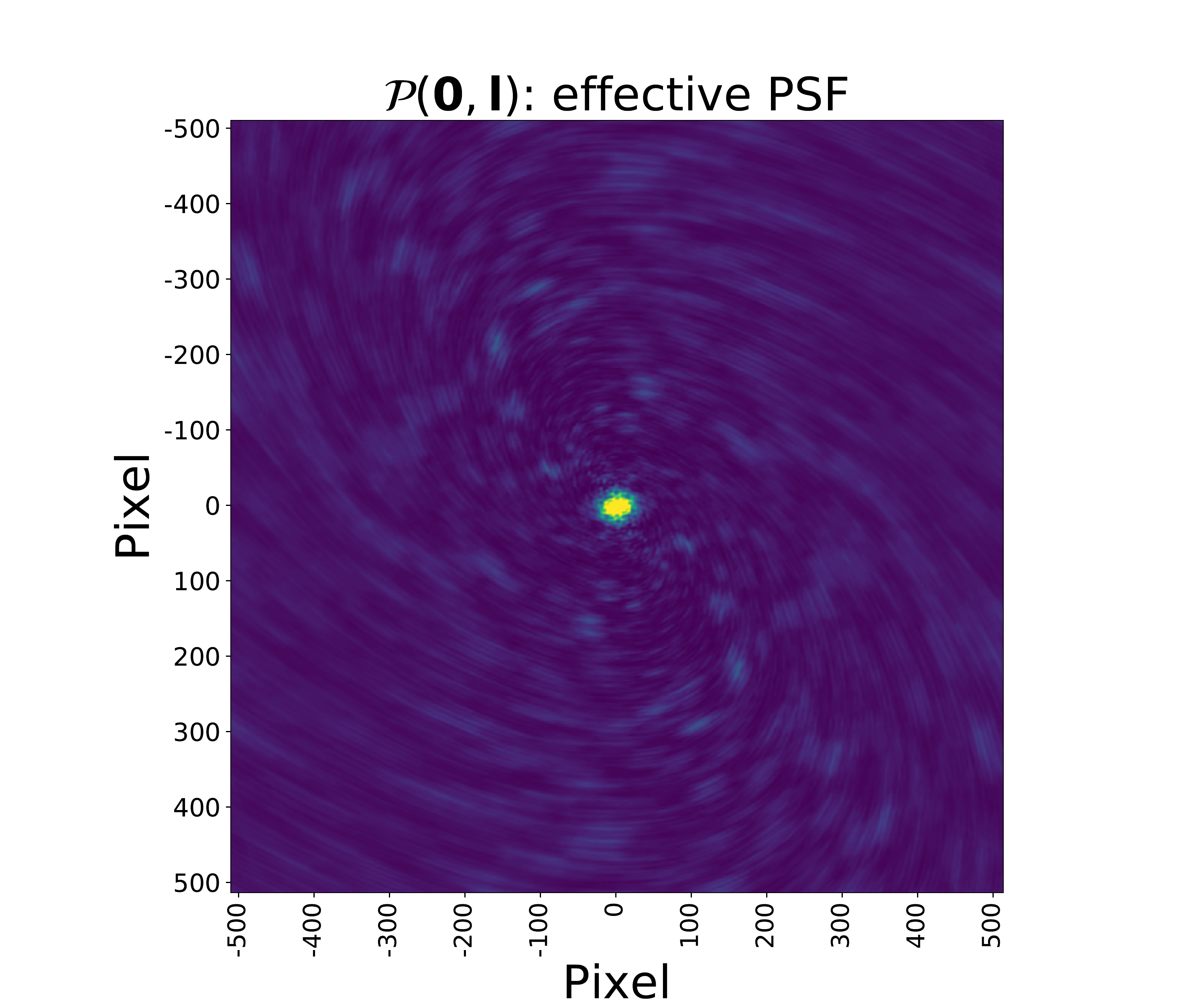}\hspace{-1.cm}
                    \includegraphics[width=.35\linewidth]{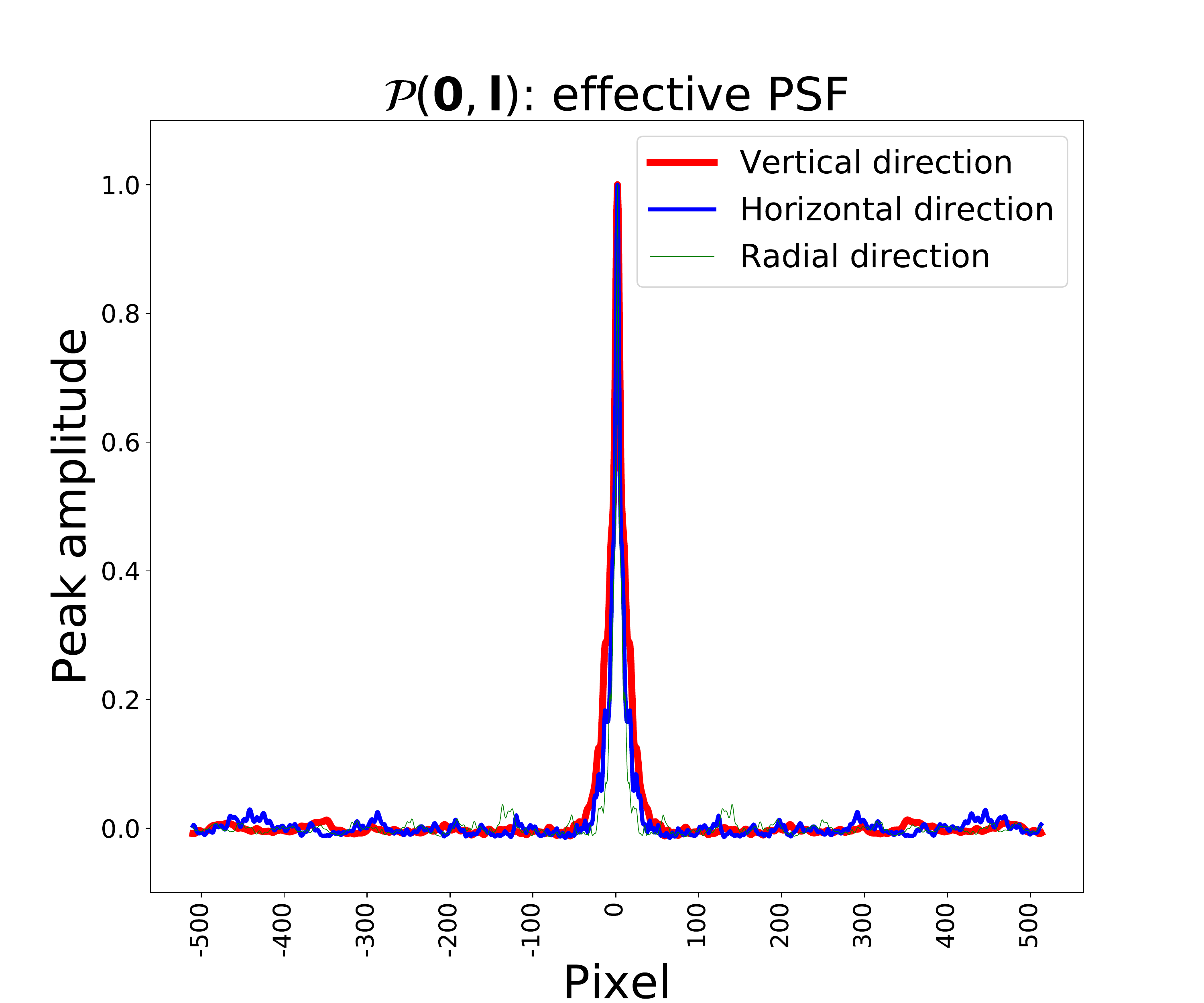}\\ 
\caption{The sampling function of the MeerKAT telescope at 1.4 GHz (\ATMB{left-panel}) for a simulated observation of 2 hrs synthesis time with 1 s integration time and 6 MHz total bandwidth, the points are where data are measured for the different baselines. The effective PSF (\ATMB{middle-panel}) and cross-sections (\ATMB{right-panel}) is the Fast  Fourier transform of the sampling function.  The sidelobes in the effective PSF show that the sampling function relies on a discretised and bandlimited space.}
\label{figpsf}
\end{figure*}

Figure~\ref{figpsf} shows a natural weighted sampling function (\ATMB{left-panel}) and the resulting effective PSF (\ATMB{middle-panel} in 2D and \ATMB{right-panel} cross-sections) of the MeerKAT telescope at 1.4 GHz for a simulated observation of $2$ hrs synthesis time with $1$ s integration time and $6$ MHz total bandwidth channelised into $120$ channels of width $50$ kHz.
Each ellipse represents the points where data are measured on the different baselines. The depicted effective PSF is the result of the Fast  Fourier Transform (FFT) of the natural weighted sampling function. 
The sidelobes in the effective PSF show that the sampling function relies on a discretised and \ATML{bandlimited} space with missing data points. 

As shown in Eq.~\ref{eq:xxPe}, the PSF remains invariant if we should be inverting only the weighted sampling function of an interferometer \ATM{array}. Evidence suggests that there \ATM{is a distortion distribution} different from sampling and weighting each visibility that \ATM{makes the PSF} position-dependent. 
\subsection{Describing the distortion distribution}
\newcommand{\rr}{\eta}
\newcommand{\hh}{\mathcal{H}}

Let us reconsider the van Cittert-Zernike theorem in Eq.~\ref{4.2.1} for a single source at $\bmath{l}_0$  with unity flux. 
For this single source the true visibility $\rr$ is the  Fourier transform  of the \ATM{ideal point source}: 
\begin{alignat}{2}
 \rr(\bmath{u}_{})=&\mathcal{F} \ATM{\{}\delta_{\bmath{l}_0}\ATM{\}}, \label{eq:fullskyangular}
\end{alignat}
\ATMB{where the ideal point source, $\delta_{\bmath{l}_0}=\delta(\bmath{l}-\bmath{l}_0)$ is a shifted delta function.}
In this case, another way to look at Eq.~\ref{sect:2x} is to consider the PSF of a single source as 
a sum of weighted PSFs where \ATM{each individual PSF} is the inverse Fourier transform of the source true visibility sampled at $pqkr$, i.e:
\begin{alignat}{2}
\mathcal{P}(\bmath{l}_0, \bmath{l})&=\sum_{pqkr}^{}W_{pqkr}\mathcal{F}^{-1}\ATM{\{}\rr_{pqkr}\ATM{\}},\label{sect:2xxxx} 
\end{alignat}
where $\rr_{pqkr}$ is the sampled version of $\rr$ at $pqkr$; i.e.:
\begin{alignat}{2}
 \rr_{pqkr}=\delta_{pqkr}\rr.
\end{alignat}

However, in reality an interferometer \ATM{array} is non-ideal, in the sense  that a measurement is the averaged visibility over 
some time and frequency \emph{sampling intervals} $\Delta t,\Delta \nu$, which  in both time-frequency directions we  denote as the \emph{sampling bin:}
\begin{equation}
\Btf_{kr} = \bigg [ t_k-\frac{\Delta t}{2},t_k+\frac{\Delta t}{2} \bigg ]
\times
\bigg [ \nu_r-\frac{\Delta\nu}{2},\nu_r+\frac{\Delta\nu}{2} \bigg ],  \label{eq:chap5resamplingbin}
\end{equation}
where $k$ and $r$ represent the indices of the centre time  and frequency bins respectively.

The averaged  measurement over the sampling bin can be represented by the integral:
\begin{alignat}{2}
\rrm_{pqkr} =& \frac{1}{\Delta t \Delta \nu} 
\iint\limits_{\Btf_{kr}}
\rr(\bmath{u}_{pq}(t,\nu))\Rd\nu \Rd t.
\label{eq:chapter5conti1}
\end{alignat}
If $\Pi^{t\nu}$ is a normalised 2D boxcar window 
then Eq.~(\ref{eq:chapter5conti1}) can be reformulated as:
 \begin{alignat}{2}
 \begin{split}
\rrm_{pqkr} &=\iint\limits_{\infty}\Pi^{t\nu}(t-t_k, \nu-\nu_r)\rr_{pq}(t,\nu)\Rd \nu\Rd t.
\label{eq:chapter5conti2}
\end{split}
\end{alignat}
If baseline-dependent window function (BDWF, ~\citet{atemkeng2016}) or baseline-dependent averaging and windowing (BDAWF, ~\citet{atemkeng2018baseline})  are employed to minimise the distortion  effects (mostly in amplitude lost) then the normalised 2D 
boxcar window is replaced by a BDWF or BDAWF, $\XX_{pqkr}$ in the $t\nu$-space:
 \begin{alignat}{2}
 \begin{split}
\rrm_{pqkr} &=\iint\limits_{\infty}\XX_{pqkr}(t-t_k, \nu-\nu_r)\rr_{pq}(t,\nu)\Rd \nu\Rd t,
\label{eq:chapter5conti2}
\end{split}
\end{alignat}
which can be expressed as a convolution in $t\nu$-space:
 \begin{alignat}{2}
 \begin{split}
\rrm_{pqkr} &=[\XX_{pqkr}\circ\rr_{pq}](t_k,\nu_r).
\label{eq:chapter5contffibb2}
\end{split}
\end{alignat}
\ATMB{The notation $[\cdot](\cdot)$ implies that the script protected by the square brackets is a function of the script in the regular brackets.} 
Likewise, Eq.~\ref{eq:chapter5contffibb2} can be written in $uv$-space:
 \begin{alignat}{2}
 \begin{split}
\rrm_{pqkr} &=[\XX_{pqkr}\circ\rr_{pq}](\bmath{u}_{pq}(t_k,\nu_r))\\
	    &=\delta_{pqkr}[\XX_{pqkr}\circ\rr_{}].
\label{eq:chapter5cdontffi2}
\end{split}
\end{alignat}
In the ideal case where there are no instrumental effects and other corruptions like distortion $\rrm_{pqkr}\equiv\rr_{pqkr}$. \ATM{The} latter remains possible only if the sampling bin is \ATMB{sufficiently small} to avoid any distortion in the signal \ATM{which is impractical because it leads} to a very large amount of data. For example, if $\mathsf{B}_{\mathrm{max}}$ is the largest sampling bin 
for which the data can be averaged without any distortion in the entire \ATM{field of view} (including the edges), we have:
\begin{equation}
\begin{cases}
\Btf_{kr}\leq \mathsf{B}_{\mathrm{max}}\\
\rrm_{pqkr}-\rr_{pqkr}\sim 0
\end{cases} 
\end{equation}
\ATM{because $\rrm_{pqkr}\equiv\rr_{pqkr}$}.
 For significant or aggressive data compression purposes, we have:
 \begin{equation}
    \begin{cases}
  \Btf_{kr}> \mathsf{B}_{\mathrm{max}} \\
   \rrm_{pqkr}-\rr_{pqkr}=e_{pqkr}\neq 0,\label{aprexxx}
\end{cases} 
    \end{equation}
\ATM{\ATMB{where the error $e_{pqkr}$ can  explicitly be written as:} 
\begin{alignat}{2}
 e_{pqkr}&=\delta_{pqkr}[\XX_{pqkr}\circ\rr_{}]-\delta_{pqkr}\rr_{}\\
	 &= \delta_{pqkr}\big[(\XX_{pqkr}\circ\mathcal{F} \ATM{\{}\delta_{\bmath{l}_0}\ATM{\}})-\mathcal{F} \ATM{\{}\delta_{\bmath{l}_0}\ATM{\}}\big],\label{eq27}
\end{alignat}
which becomes  bigger with increasing  $\Btf_{kr}$ and $\bmath{l}_0$} \ATMB{since $\XX_{pqkr}$ deviates further from $\delta_{pqkr}$}:
this is the main cause for the distortion of the PSF because $\rr_{pqkr}$ is  replaced  in  Eq.~\ref{sect:2xxxx} by $\rrm_{pqkr}$:
\begin{alignat}{2}
\mathcal{P}_d(\bmath{l}_0, \bmath{l})&=\sum_{pqkr}^{}W_{pqkr}\mathcal{F}^{-1}\ATM{\{}\rrm_{pqkr}\ATM{\}}\\
	       &=\sum_{pqkr}^{}W_{pqkr}\Big(\mathcal{P}_{pqkr}\circ\mathcal{F}^{-1} \ATM{\{}\XX_{pqkr}\ATM{\}}\mathcal{F}^{-1}\ATM{\{}\rr_{}\ATM{\}}\Big),\label{eq:de}
\end{alignat}
which becomes position-dependent.
Knowing that $\mathcal{F}^{-1}\ATM{\{}\rr_{}\ATM{\}}=\delta_{\bmath{l}_0}$ (see Eq.~\ref{eq:fullskyangular}), we deduce that the distortion distribution is given by:
\begin{alignat}{2}
 \mathcal{D}_{pqkr}=\mathcal{F}^{-1}\ATM{\{}\XX_{pqkr}\ATM{\}}.
\end{alignat}
Here, $\mathcal{D}_{pqkr}= \mathcal{D}_{pqkr}(\bmath{l}_0)$ describes the  sampled and truncated image-plane baseline-dependent distortion distribution.  

\begin{figure*}
\centering
  \includegraphics[width=.8\linewidth]{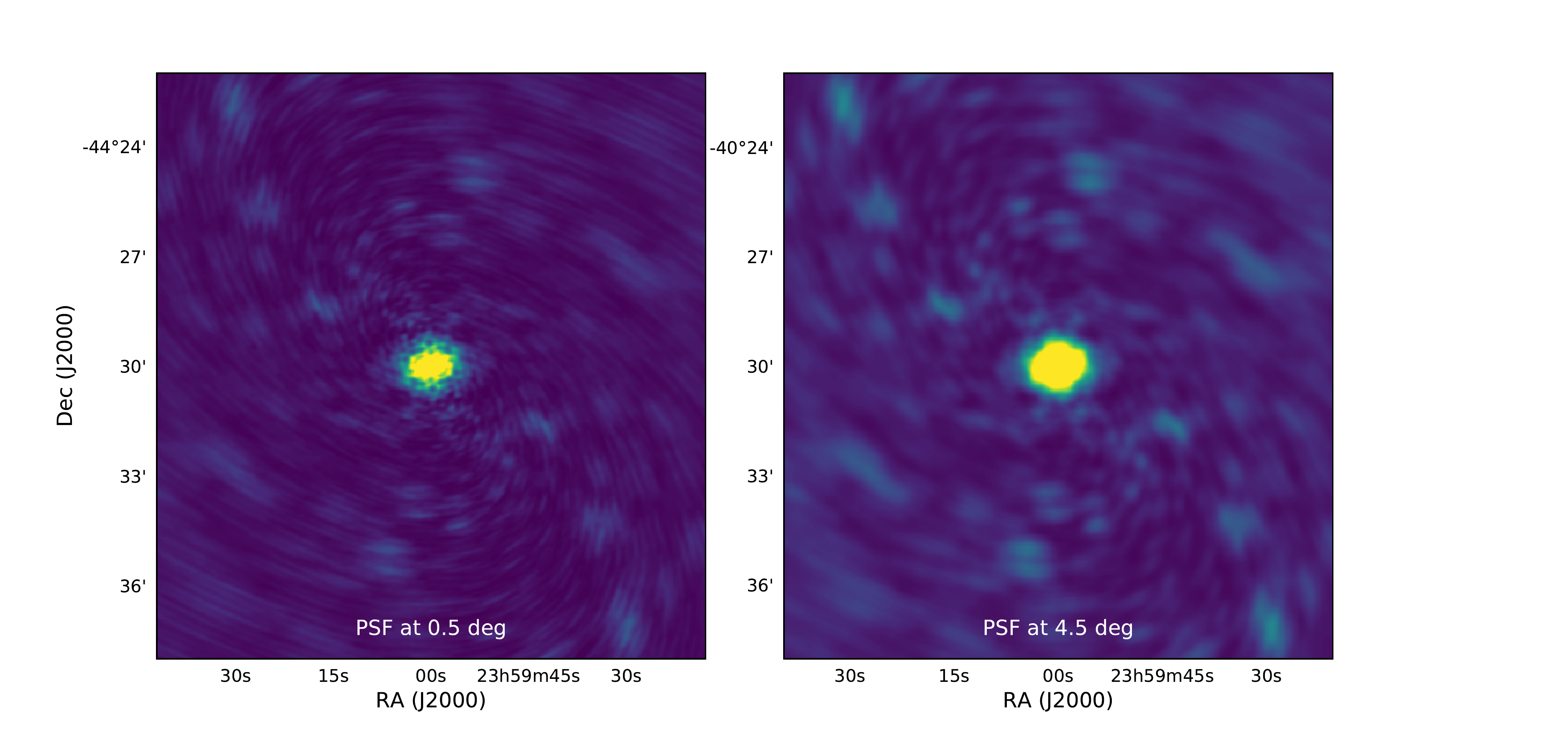}
\caption{The normalised position-dependent PSFs for a source at $0.5$ deg (left) and $4.5$ deg  (right) for a simulated observation of the MeerKAT telescope at $1.4$ GHz.  The data is sampled at $1$ s and $50$ kHz during 2 hrs with a total bandwidth of $6$ MHz and resampled by averaging
$20$ samples  across the time and  $40$ samples across the frequency direction making a total compression factor of $\text{CF} = 20 \times 40$. }
\label{fig:psf}
\end{figure*}

\begin{figure}
\centering
    \includegraphics[width=.8\linewidth]{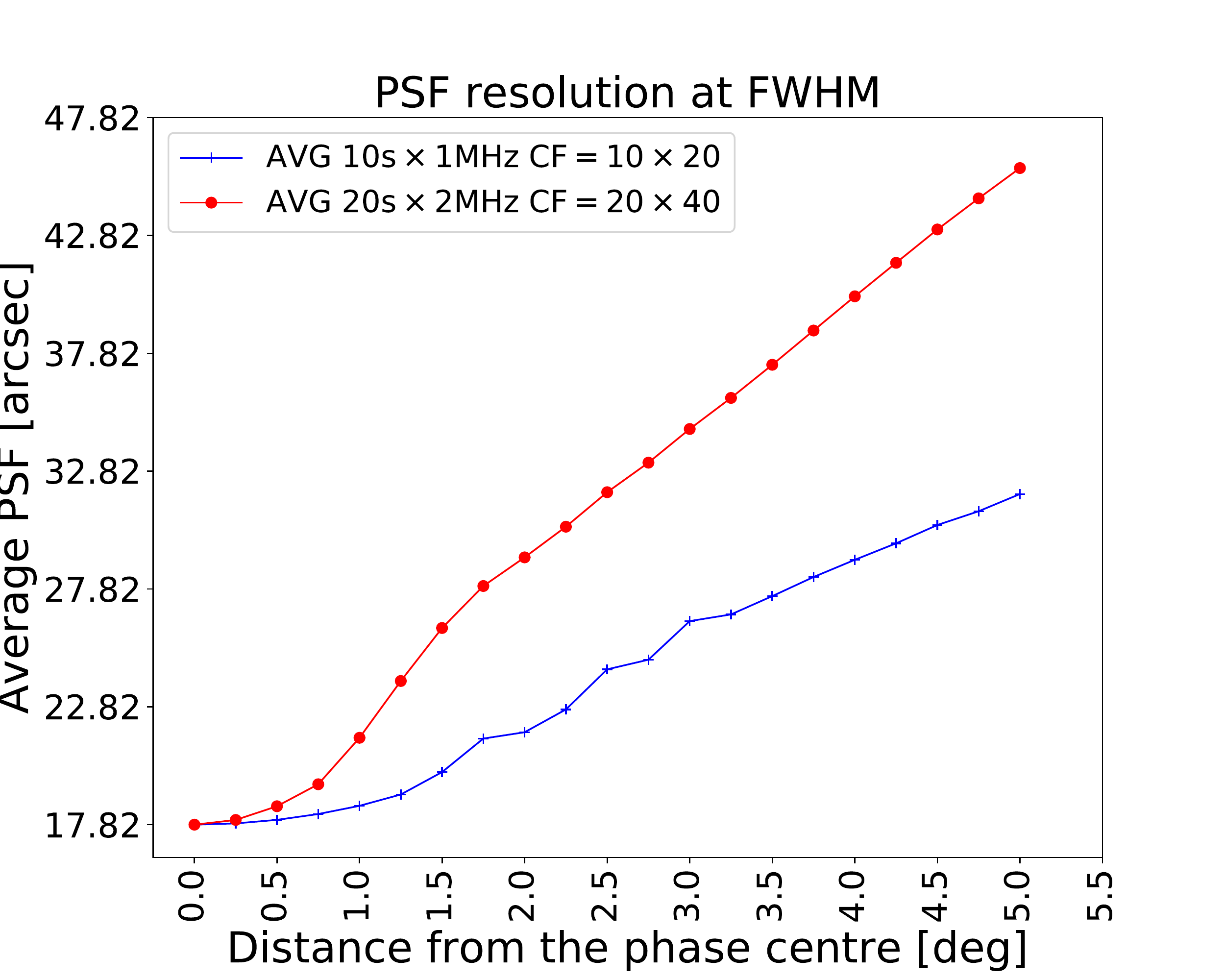}%
\caption{Simulation of the averaged FWHM resolution of the PSF in the radial and tangential direction as a function of distance from the phase centre for the  MeerKAT at 1.4 GHz after 2 hrs using a total bandwidth of 6 MHz.}
\label{figpsfVal}
\end{figure}

Figure~\ref{fig:psf} shows two natural weighted \ATM{position-dependent} PSFs at $0.5$ deg and $4.5$ deg away from the phase centre of the MeerKAT telescope at $1.4$ GHz. \ATMB{The simulated data described in Section~\ref{sect2.2} is aggressively averaged} for  compression reasons, e.g., resampled by averaging
$20$ samples  across the time direction and  $40$ samples across the frequency direction making a total compression factor of $\text{CF} = 20 \times 40$ and the averaged integration  becomes $20 ~\text{s}$  and $2 ~\text{MHz}$ in time and in frequency respectively. The latter are \ATMB{denoted by} the notation $\text{AVG}~20\text{s} \times 2 \text{MHz CF}=20\times 40$. 
We see that by doing this aggressive compression, we distort the  \ATM{PSF} at each source differently. \ATML{The \ATMB{size} of these PSFs becomes a function of position in the sky with wider \ATMB{size}  as we move from the phase centre of the observation.} 
Figure~\ref{figpsfVal} \ATMB{quantifies the size} of the position-dependent PSFs measured at \ATMB{their FWHM} as a function of 
distance from the phase centre for two compression factors; i.e. $\text{AVG}~10\text{s}\times1\text{MHz CF}=10\times 20$ and $\text{AVG}~20\text{s} \times 2 \text{MHz CF}=20\times 40$. This confirms that the \ATMB{size} of the position-dependent 
PSFs also depends on the rate of compression and therefore the sampling bin. In this work, we do not measure the degree of the position-dependent PSFs amplitude attenuation because the attenuated amplitude quantifies the degree of the smeared source amplitude which is explained in~\citet{atemkeng2016}. \ATM{We have described $\mathcal{D}_{pqkr}$ analytically. In Section~\ref{fatsderivationspeudopsfs} we present two algorithms to compute $\mathcal{P}_d$ quicker by approximation.} 

\section{Fast algorithms to approximate the position-dependent PSFs}
\label{fatsderivationspeudopsfs}

In this section, we present two analytical frameworks that can be used to approximate the position-dependent PSFs with fewer computational requirements. 
We describe the main difference between the two analytical frameworks and show that they are computationally cheaper when we compare with the computing cost using a brute-force approach.

\subsection{Method 1: $uv$-plane approximation}
\label{uvdomainapproximation}
Let us reconsider Eq.~\ref{eq:de} where $\mathcal{F}^{-1}\ATM{\{}\rr_{}\ATM{\}}$ has been substituted by $\delta_{\bmath{l}_0}$:
\begin{alignat}{2}
\mathcal{P}_{d}(\bmath{l}_0, \bmath{l})&=\sum_{pqkr}^{}W_{pqkr}\Big(\mathcal{P}_{pqkr}\circ\mathcal{F}^{-1} \ATM{\{}\XX_{pqkr}\ATM{\}}\delta_{\bmath{l}_0}\Big)\\
				       &=\ATMB{\sum_{pqkr}^{}W_{pqkr}\Big(\mathcal{P}_{pqkr}\circ\mathcal{F}^{-1} \ATM{\{}\XX_{pqkr}\ATM{\}}\Big),}\label{eq:de11}
\end{alignat}
 \ATMB{where $\delta_{\bmath{l}_0}\sim 1$.} For a baseline $pq$, the  image-plane distortion distribution $ \mathcal{D}_{pqkr}=\mathcal{F}^{-1}\ATM{\{}\XX_{pqkr}\ATM{\}}$   is the sampled and truncated  version of the true image-plane distorted distribution $\mathcal{D}$. The latter is related to its sampled and truncated version within the sampling bin as:
\begin{alignat}{2}
\ATMB{[}\mathcal{F}\ATM{\{}\mathcal{D}_{pqkr}\ATM{\}}\ATMB{]}(\bmath{u}_{}) &= \ATMB{[}\mathcal{F}\ATM{\{}\mathcal{D}\ATM{\}}\ATMB{]}(\bmath{u}_{}-\bmath{u}_{pqkr})\\
				 &=\delta_{}(\bmath{u}_{}-\bmath{u}_{pqkr})\circ \ATMB{[}\mathcal{F}\ATM{\{}\mathcal{D}\ATM{\}}\ATMB{]}(\bmath{u}_{})\label{eq:meho1:3}.
\end{alignat}
Inverting the relation in Eq.~\ref{eq:meho1:3} we arrive at:
\begin{alignat}{2}
\mathcal{D}_{pqkr}(\bmath{l}_{0}) &=\ee^{-2i\pi \bmath{u}_{pq}(t_k, \nu_r)\bmath{l}_0}\mathcal{D}(\bmath{l}_{0}).
\end{alignat}
The true image-plane distortion distribution $\mathcal{D}(\bmath{l}_0)$ is of critical importance in this work, so it warrants a detailed explanation. In the case of an ideal scenario $\mathcal{D}(\bmath{l}_0)$ is a continuous and untruncated function that measures the true  distortion at $\bmath{l}_0$. 
\ATMB{The distortion and attenuation are baseline-dependent because an interferometer \ATM{array} introduces sampling and truncation biases which are different at each baseline that is part of the interferometer \ATM{array}}.
In practical scenarios $\mathcal{D}(\bmath{l}_0)$ is approximated by the accumulation effect of all these sampled and truncated baseline-dependent individual image-plane distortion distribution $\mathcal{D}_{pqkr}(\bmath{l}_0)$.
The continuous measurement $\mathcal{D}(\bmath{l}_{0})$ can be
approximated for each baseline in terms of the phase changes at each averaged time $t_k$ and frequency $\nu_r$  as:
\begin{alignat}{2}
\widehat{\mathcal{D}}(\bmath{l}_{0}) &=\mathcal{D}\bigg(\frac{\Delta \Psi_{pq}}{2}, \frac{\Delta \Phi_{pq}}{2}\bigg).
\end{alignat}
$\mathcal{D}_{pqkr}(\bmath{l}_{0})$ and $\mathcal{P}_{d}(\bmath{l}_0, \bmath{l})$ become also an approximation:
\begin{alignat}{2}
\widehat{\mathcal{D}}_{pqkr}(\bmath{l}_{0}) &=\ee^{-2i\pi \bmath{u}_{pq}(t_k, \nu_r)\bmath{l}_0}\widehat{\mathcal{D}}(\bmath{l}_{0})\label{visAproxxxD}\\
\widehat{\mathcal{P}}_{d}(\bmath{l}_0, \bmath{l})&=\sum_{pqkr}^{}W_{pqkr}\Big(\mathcal{P}_{pqkr}\circ\widehat{\mathcal{D}}_{pqkr}\Big),\label{eq:de11xxx}
\end{alignat}
where $\Delta \Psi_{pq}$ and $\Delta \Phi_{pq}$ are the phase difference in time and frequency respectively.
We define these phases as follows:
\begin{alignat}{2}
\begin{split}
\Delta \Psi_{pq}  =&2\pi \bigg(\big(u_{pq}(t_s,\nu_r)-u_{pq}(t_e,\nu_r)\big)l_0 \\
                  +& \big(v_{pq}(t_s,\nu_l)-v_{pq}(t_e,\nu_r)\big)m_0\bigg)\\
	      =&2\pi\Delta \bmath{u}_{pq}(t,\nu_r)\bmath{l}_0
\end{split}\\
\begin{split}
\Delta \Phi_{pq}  =&2\pi\bigg((u_{pq}(t_k,\nu_s)-u_{pq}(t_k,\nu_e))l_0 \\
		+& (v_{pq}(t_k,\nu_s)-v_{pq}(t_k,\nu_e))m_0 \bigg)\\
	     =&2\pi\Delta \bmath{u}_{pq}(t_k,\nu)\bmath{l}_0,
\end{split}	    
\end{alignat}
where 
 \begin{equation}
    \begin{cases}
t_s=t_k-\frac{\Delta t}{2}\\
t_e=t_k+\frac{\Delta t}{2}\\
\nu_s=\nu_r-\frac{\Delta\nu}{2}\\
\nu_e=\nu_r+\frac{\Delta\nu}{2} 
\end{cases} 
    \end{equation}
\ATMB{are} the  
starting time, ending time, starting frequency and ending frequency of the sampling bin respectively. 
Note from this approximation that  all the  visibilities measured at times $t_{i\neq k}$ and frequencies $\nu_{j\neq r}$ are discarded during the approximation.  \ATMB{Only the visibility: $\ee^{-2i\pi \bmath{u}_{pq}(t_k, \nu_r)\bmath{l}_0}$}
is used for the approximation (see Eq.~\ref{visAproxxxD}), which results in cheaper computations. 

If $\XX$ is treated as a 2D boxcar window function and knowing that $\mathcal{D}_{pqkr}=\mathcal{F}^{-1}\ATM{\{} \mathcal{X}_{pqkr}\ATM{\}}$ then:
\begin{alignat}{2}
\widehat{\mathcal{D}}(\bmath{l}_{0}) &= \mathcal{D}\big(\frac{\Delta \Psi_{pq}}{2}, \frac{\Delta \Phi_{pq}}{2}\big)\\
&=\mathrm{sinc}\big(\frac{\Delta \Psi_{pq}}{2}\big) \mathrm{sinc}\big(\frac{\Delta \Phi_{pq}}{2}\big),
\end{alignat}
where the 2D $\mathrm{sinc}$ comes from the 2D inverse
Fourier transform of a boxcar window function. 
\subsection{Method 2: image-plane approximation}
\label{imageplaneapproximation}
\ATMB{Suppose that all the baselines have the same distortion distribution in the image-plane which is $\mathcal{D}(\bmath{l}_0)$:
\begin{alignat}{2}
\mathcal{D}_{pqkr}(\bmath{l}_{0}) &\sim\mathcal{D}(\bmath{l}_{0}).
\end{alignat}
\ATMB{Eq.~\ref{eq:de11}} can be treated as}  a convolution between the effective PSF and the true sky distorted distribution $\mathcal{D}(\bmath{l}_0)$:
\begin{alignat}{2}
 \mathcal{P}_{d}(\bmath{l}_0, \bmath{l})  =& \PP(\textbf{0}, \bmath{l})\circ\mathcal{D}(\bmath{l}_0).\label{eq:meth2:1}
 \end{alignat}
The effective PSF $\PP(\textbf{0}, \bmath{l})$ is known while $\mathcal{D}(\bmath{l}_0)$ is unknown. In the following paragraphs, we discuss an algorithm to find \ATML{an  approximation} for $\mathcal{D}(\bmath{l}_0)$.

In the time domain, assume the baselines  trace out a perfect circle in the $uv$-plane as 
if they were  east-west baselines (i.e. without \EDIT{a} $v$-offset in the ellipses) and 
observing a source at the Zenith. The baseline\EDIT{,} which samples the bin  at $(u-u_0,v-v_0)$\EDIT{,} 
 has for vector $\bmath{u}_0=(u_0,v_0)$.  This implies that 
a $uv$-track in time is a perfect circle with radius $||\bmath{u}_0||$  and
 angular velocity of $\omega_E$: 
 \ATM{\begin{alignat}{2}
  u(t) &=\frac{\nu}{c}\omega_E||\bmath{u}_0||\frac{\partial t}{\partial \theta}\cos\big(\theta(t)\big)\label{eq:upsf}\\
  v(t) &=\frac{\nu}{c}\omega_E||\bmath{u}_0||\frac{\partial t}{\partial \theta}\sin\big(\theta(t)\big),\label{eq:vpsf}
 \end{alignat}}
where \ATM{$\theta(t)=\arctan(u_0(t)/v_0(t))$} is the angle of orientation. The fringes rotation speed in time is then given by
 the partial derivative of Eq.~(\ref{eq:upsf}) and (\ref{eq:vpsf})\EDIT{:}
 \ATM{\begin{alignat}{2}
  \frac{\partial u}{\partial t} &=-\frac{\nu}{c}\omega_E||\bmath{u}_0||\sin\big(\theta(t)\big)\label{eq:upsf2}\\
  \frac{\partial v}{\partial t} &=\frac{\nu}{c}\omega_E||\bmath{u}_0||\cos\big(\theta(t)\big).\label{eq:vpsf2}
 \end{alignat}}
The  \ATML{approximation} of the phase difference in time is then derived as\EDIT{:}
 \begin{alignat}{2}
  \Delta \Psi  &\approx 2\pi\Delta t \frac{\partial \bmath{u}}{\ATM{\partial t}}\bmath{l}_0\\
		&\approx 2\pi(\frac{\partial u}{\ATM{\partial t}}l_0+\frac{\partial v}{\ATM{\partial t}}m_0)\Delta t.
 \end{alignat}
In the frequency domain, decorrelation can be characterised by the product of the fractional 
bandwidth $\Delta \nu /\nu$ with the source distance $||\bmath{l}_0||$ from the phase centre   relative to the baseline length $||\bmath{u}_0||$.
An \ATML{approximation} of the phase difference in frequency is given by\EDIT{:}
 \begin{alignat}{2}
  \Delta \Phi  &\approx   2\pi\frac{\Delta \nu}{\nu}||\bmath{l}_0||\times||\bmath{u}_0||\\
   &\approx 2\pi\frac{\Delta \nu}{\nu}\Big((l_0^2+m_0^2)(u_0^2+v_0^2)\Big)^{\frac{1}{2}}.
 \end{alignat}
\ATM{The true} sky distorted distribution function $\mathcal{D}(\bmath{l}_0)$ is 
approximated as\EDIT{:}
\begin{alignat}{2}
\widehat{\mathcal{D}}(\bmath{l}_0) = \mathcal{D}\bigg(\frac{\Delta \Psi}{2}, \frac{\Delta \Phi}{2}\bigg).
\end{alignat} 
The above processes for approximating $\mathcal{D}(\bmath{l}_0)$ is \ATMB{summarised in Algorithm~\ref{alg:decribtion} where $\mathcal{D}$ is represented by} a pixelarised matrix $\bmath{\mathcal{D}}$.

Because it is often efficient to use the Fourier transform to compute a convolution, once $\mathcal{D}(\bmath{l}_0)$ has been approximated, the \ATML{approximation} of Eq.~\ref{eq:meth2:1} is rewritten as:
 \begin{alignat}{2}
 \widehat{\mathcal{P}}_{d}(\bmath{l}_0, \bmath{l})=&\mathcal{F}^{-1}\Big\{[\mathcal{F}\ATM{\{}\PP\ATM{\}}\mathcal{F}\ATM{\{}\widehat{\mathcal{D}}\ATM{\}}](\bmath{u})\Big\},\label{eq:imageplaneaproxiphase}
\end{alignat}
which is computationally cheaper given that $\PP$ is computed once for all $\bmath{l}_0$. 
\begin{algorithm}
\centering
\begin{algorithmic}[1]
\Procedure{Approximation of $\bmath{\mathcal{D}}$.}{}
\State $\Delta u:=\frac{1}{N_l\Delta m}$, $\Delta v:=\frac{1}{N_m\Delta l}$, $u_0:=\frac{1-N_l}{2}\Delta u$ 
\For{ $i$ from $1$ to $N_l$ }
\State $v_0:=\frac{1-N_m}{2} \Delta v$
\For{ $j$ from $1$ to $N_m$ }
\State $uv_0 = \omega_E\sqrt{u_0^2+v_0^2}$
\State $\theta := \arctan(u_0/v_0)$
\State $\dot{u} := - uv_0\sin \theta$
\State $\dot{v} :=  uv_0\cos \theta$
\State $\Delta \Psi_{}:= \frac{\nu}{c}2\pi(\dot{u}l_i+ \dot{v}m_j)\Delta t$
\State $\Delta \Phi_{}:= 2\pi  \frac{\Delta \nu}{\nu}\frac{uv_0}{\omega_E}\sqrt{l_i^2+m_j^2}$
\State $\widehat{\bmath{\mathcal{D}}}_{ij}:=\mathcal{D}(\Delta \Psi/2, \Delta \Phi/2)$ 
\State $v_{0}:=v_{0}+\frac{N_m-1}{N_m}\Delta v$
\EndFor
\State $u_{0}:=u_{0}+\frac{N_l-1}{N_l}\Delta u$
\EndFor
\EndProcedure
\end{algorithmic}
\caption{The $uv$-plane is a discretized measurement of dimension
$N_l\Delta u \times N_m\Delta v$,
where $N_l N_m$ is the total number of pixels. The discretized bins are separated
by the amount of   $\Delta u$ and $\Delta v$ in the $u$ and $v$ direction respectively. In the image-plane, the pixels are separated by the amount of $\Delta l$ and $\Delta m$ in the $l$ and $m$ direction respectively.
}\label{alg:decribtion}
\end{algorithm}
\subsection{Computational costs} 
\label{computationcostPSF}
We use the  \citet{hogbom1974aperture} CLEAN algorithm approach to showcase the computational complexity of any CLEAN based algorithm
using position-dependent PSF\ATM{s} during the deconvolution iterations. Note that the complexity only relies on the steps that involve the position-dependent PSFs computation rather than the entire CLEAN algorithm. 
The  H\"{o}gbom CLEAN algorithm using the position-dependent PSFs follows the procedure in Algorithm~\ref{alg:hogbom}.
At each of the iteration in Algorithm~\ref{alg:hogbom} a position-dependent  PSF is computed (see line $\ref{ligne5}$) \ATMB{before being properly normalised} with the peak pixel value of the dirty image then 
subtracted from the dirty image (line~\ref{ligne6} of the algorithm). \ATML{The number of visibilities is large for these big data instruments, therefore, the FFT  is used to compute the  Fourier transform in Eqs.~\ref{eq:de11xxx} and \ref{eq:imageplaneaproxiphase}. In order to compute
the FFT, the visibility data is interpolated onto a regular grid. In all the computational complexities performed in this section, we assume that the FFT is not uniform; the computational complexity for the gridding is not dissociable from the computational complexity of the FFT itself as shown in \citep{cooley1965algorithm, smith2017reconsidering, ye2020optimal}. The computational complexity  $C_{\PP}$ of the  non-uniform FFT for computing the effective PSF scales as:}
\begin{alignat}{2}
C_{\PP}\sim\OO   \Big( N_{pq} N_{t}^{} N_{\nu}^{}\log_{2}(N_lN_m)\Big),\label{psf-phase-centre1}
\end{alignat}
where $N_lN_m$ is the number of pixels in the \ATMB{dirty image} with $N_l$ and $N_m$ the number of pixels in $l$ and $m$ direction respectively. Here, $N_{pq}$ is the number of baselines, $N_{t}^{}$ and  $N_{\nu}^{}$ are the number time and frequency bins respectively. 
The product $N_{pq} N_{t}^{} N_{\nu}^{}$ is the total number of sampled visibilities and it predicts the time taken to evaluate the fringe induced by each baseline, multiplied by the source amplitude and followed by the summation over all the  \ATMB{visibilities}.  
For big-data interferometer \ATM{arrays} even when using the 2D van Cittert-Zernike theorem for an \ATML{approximation} of the  wide-field (i.e. very large $N_l  N_m$), we have:
\begin{equation}
 N_l  N_m \ll N_{pq} N_{t}^{} N_{\nu}^{}.
\end{equation}

\paragraph*{Complexity to compute all the $\PP_{d}(\bmath{l}_0, \bmath{l})$ in Algorithm~\ref{alg:hogbom} using brute-force:}
Assume   that $N_\mathrm{src}$ is the number of sources in $\IID$ or the number of iterations in Algorithm~\ref{alg:hogbom} before the peak in $\IID$ hits the noise level. The complexity, $C_{\PP_d}^{\mathrm{bf}}$ to evaluate all the $N_\mathrm{src}$ 
position-dependent PSFs  by brute-force
will scale as:
\begin{alignat}{2}
C_{\PP_d}^{bf}&\sim \OO  \Big(N_\mathrm{src}C_{\PP}\Big)\\
&\sim \OO  \Big(N_\mathrm{src}N_{pq} N_{t}^{} N_{\nu}^{}\log_{2}(N_lN_m)\Big).\label{brute:comp}
\end{alignat}
In the worst case where each pixel in $\IID$ is a source, we have $N_\mathrm{src}\sim N_lN_m$  and the brute-force predicted cost to evaluate the position-dependent PSFs now scale as:
\begin{alignat}{2}
C_{\PP_d}^{bf}\sim  \OO  \Big(N_lN_m N_{pq} N_{t}^{} N_{\nu}^{} \log_{2}(N_lN_m)\Big),\label{complexity:iter}
\end{alignat}
which scales very poorly for big-data interferometer \ATM{arrays} and wide-field imaging.

\paragraph*{Complexity using  method 1:}
The $uv$-plane approximation for the position-dependent PSFs uses only the \ATMB{visibilities whose phases are the phase gradients}
and discards other visibilities. 
Thus, if $N_{pq} N_{t}^{} N_{\nu}^{}$  is the total number of visibilities\EDIT{,} then only a few visibility
samples are used for the $uv$-plane approximation. In other words, if $n_t$ is the number of visibilities to average in time and  $n_{\nu}$ the number of visibilities to average in frequency 
then 
\begin{alignat}{2}
 N_{pq} N_{k} N_{r}=N_{pq}\frac{N_{t}^{}}{n_t}\frac{N_{\nu}^{}}{n_{\nu}}
\end{alignat}
 will be the number of phase gradients used in the approximation. 
The complexity to approximate a unique position-dependent PSF  in $uv$-plane scales as:
\begin{alignat}{2}
 C_{\widehat{\PP}_{d}}&\sim \OO  \Big( N_{pq}N_{k} N_{r}\log_{2}(N_lN_m)\Big).
\end{alignat}
The phase gradient is
different for each source, which is problematic as it emphasises that a phase gradient must be computed at each iteration, therefore, the complexity will increase for all the position-dependent PSFs  by a factor of $N_\mathrm{src}$. In the worst case
$N_\mathrm{src}\sim N_lN_m$ we have:
\begin{alignat}{2}
C_{\widehat{\PP}_{d}}^{uv}  &\sim \OO \Big(N_lN_m C_{\widehat{\PP}_{d}}\Big)\\
				  &\sim \OO  \Big( \ATM{N_lN_m  N_{pq}N_{k} N_{r}}\log_{2}(N_lN_m)\Big),\label{complexitypsf:iteruv}
\end{alignat}
which is much cheaper than the brute-force approach in Eq.~\ref{complexity:iter}. 
\algdef{SE}[DOWHILE]{Do}{doWhile}{\algorithmicdo}[1]{\algorithmicwhile\ #1}%
\begin{algorithm}
\centering
\begin{algorithmic}[1]
\Procedure{ From $\IID$ find $\IIA$ an estimate of $\II$.}{}
\State $\IIA:=0$ 
\Do
\State $\bmath{l}_0 := \argmax (\IID)$ /*$\bmath{l}_0=(i, j)$; $i$ and $j$ are the indices of the peak pixel*/.
\State Compute  $\PP_{d}(\bmath{l}_0, \bmath{l})$ /*the PSF at $\bmath{l}_0$ \label{ligne5}*/
\State $\IID:=\IID-\PP_{d}(\bmath{l}_0, \bmath{l})\circ\gamma\IID(\bmath{l}_0)$ \label{ligne6}
\State $\IIA:=\gamma \IID(\bmath{l}_0)$ /*$\gamma$ is the CLEAN gain*/
\doWhile{(peak in $\IID$  is above the noise level)}
\EndProcedure
\end{algorithmic}
\caption{H\"{o}gbom  CLEAN using position-dependent PSFs.
}\label{alg:hogbom}
\end{algorithm}
\paragraph*{Complexity using method 2:}
The image-plane approximation uses the effective PSF  to approximate all the position-dependent PSFs. 
The complexity \EDIT{of evaluating}  the effective PSF scales as:
$\OO  \big( N_{pq} N_{t}^{} N_{\nu}^{}\log_{2}(N_lN_m)\big) $ (see  Eq.~(\ref{psf-phase-centre1})). 
 Suppose that $\OO \big(\xi\big)$ is the computational
complexity to evaluate the cumulative distortion effects in Algorithm~\ref{alg:decribtion} for all sources in the \ATMB{dirty image.}  The 
image-plane approximation for the position-dependent PSFs shows the computation scaling:
\begin{alignat}{2}
 C_{\widehat{\PP}_{d}}^{lm}\sim\OO  \Big( N_{pq} N_{t}^{} N_{\nu}^{}\log_{2}(N_lN_m)\Big)+ \OO \big(\xi\big),\label{complexitypsfs:iterimageplane}
\end{alignat}
where the cost $\OO \big(\xi\big)$ can be regarded as negligible\EDIT{,} given that the evaluation of  $\widehat{\mathcal{D}}(\bmath{l}_0)$  
does not involve any exponential functions (see Section~\ref{imageplaneapproximation}). Taking the latter into account, we see that the complexity in Eq.~(\ref{complexitypsfs:iterimageplane}) is lower compared to that of the $uv$-plane approximation and that of the brute-force:
\begin{alignat}{2}
 C_{\widehat{\PP}_{d}}^{lm}<C_{\widehat{\PP}_{d}}^{uv}<C_{\PP_{d}}^{bf}.\label{computimCom}
\end{alignat}
\paragraph*{\ATM{Complexity for brute-force PSF per facet:}}
\ATM{In the case of faceting imaging, where the \ATMB{dirty image} is partitioned into facets and
each facet is deconvolved separately with the position-dependent PSF at the centre of the facet before the results of
each clean facet are merged, the computational complexity of evaluating the position-dependent PSFs by brute-force for $N_\mathrm{facet}$ facets is
\begin{alignat}{2}
C_{\mathrm{facet}}^{bf} &\sim \OO  \Big(N_{\mathrm{facet}}N_{pq} N_{t}^{} N_{\nu}^{}\log_{2}(N_lN_m)\Big),\label{comp:facte:smallsize1}
\end{alignat}
which is much better than the complexity in Eq.~\ref{complexity:iter} as  $N_{\mathrm{facet}}\ll N_lN_m$. If the far sidelobes of the position-dependent PSFs are below some given threshold, then the
size of the  position-dependent PSF per facet can be restricted to \ATMB{the size of the facet} $N_{l,\mathrm{facet}}\times N_{m, \mathrm{facet}}< N_lN_m$. In this case the
computational complexity now scales as: 
\begin{alignat}{2}
C_{\mathrm{facet}}^{bf} &\sim \OO  \Big(N_{\mathrm{facet}}N_{pq} N_{t}^{} N_{\nu}^{}\log_{2}(N_{l, \mathrm{facet}} N_{m, \mathrm{facet}})\Big),\label{comp:facte:smallsize2}
\end{alignat}
which runs much faster compared to Eq.~\ref{comp:facte:smallsize1}. Using the approximation methods described above, a faceting framework can also approximate the per facet position-dependent PSFs which will further save computations.}
All the above complexities are summarised in  Table~\ref{tab:relativeSNR}.
\begin{table*}
\begin{tabular}{ |p{7cm}||p{6cm}| }
 \hline
 \bf Methods to compute the position-dependent PSFs &  \bf Computational complexity \\
 \hline
 \hline
 Brute-force   & $\OO \big(N_lN_m N_{pq} N_{t}^{} N_{\nu}^{} \log_{2}(N_lN_m)\big)$\\
 Method 1: $uv$-plane approximation       & $\OO  \big(\ATM{N_lN_m  N_{pq}N_{k} N_{r}}\log_{2}(N_lN_m)\big)$\\
 Method 2: image-plane approximation     & $\OO  \big( N_{pq} N_{t}^{} N_{\nu}^{}\log_{2}(N_lN_m)\big)+ \OO (\xi)$\\
 \ATM{Complexity for brute-force PSF per facet} &  \ATM{$\OO  \big(N_{\mathrm{facet}}N_{pq} N_{t}^{} N_{\nu}^{}\log_{2}(N_{l, \mathrm{facet}} N_{m, \mathrm{facet}})\big)$}\\
 \hline
\end{tabular}
\caption{Corresponding computational complexities held by each method used to compute the position-dependent PSFs. The time scales for the computation  \ATMB{are shown} in term of the computational complexity.}
\label{tab:relativeSNR}
\end{table*}
\section{Simulations}
\label{simulations}
To illustrate the accuracy of the two algorithms presented in Section~\ref{fatsderivationspeudopsfs} to approximate the position-dependent PSFs,  a practical example using MeerKAT data will be presented in this section. 
We reconsider the simulated MeerKAT dataset at 1.4 GHz used in Section~\ref{sect2}. The dataset is sampled at $1$ s and $50$ kHz for 2 hrs with a total bandwidth of $6$ MHz; prepared to receive the visibilities for a single $1$ Jy point source at $0.5$ deg, $2.5$ deg, and $4.5$ deg. Each of the point sources is simulated separately then $20$  and  $40$ samples are averaged in time and in frequency respectively. This results in a resampled dataset with $20 ~\text{s}$  and $2 ~\text{MHz}$ sampling in time and in frequency respectively. 
After each of the point sources is simulated and averaged,  using the WSclean imager \citep{offringa2014wsclean} we then make a natural weighted image of size $1024\times 1024$ centered at each of the point sources. Note that the weighting scheme does not change the results of the approximation. To translate each of the simulated and resampled  datasets to the source local PSF, the WSclean performs the FFT  on the entire dataset. 
This provides us with the exact local PSF:  in the sense that apart from using convolutional kernels to avoid errors as the visibilities are non-coplanar and  gridding kernels that must satisfy the requirements of the FFT, the WSclean uses the dataset as a whole with no single visibility discarded during the imaging process.
Note that using the whole dataset without discarding any single visibility when imaging each local PSF is one of the reasons why the brute-force computation of the local PSFs is slower than our approximation. 

Our approximation approaches of the position-dependent PSFs as described in Section~\ref{fatsderivationspeudopsfs} \ATMB{introduce}  an accuracy error of:
\begin{alignat}{2}
 \mathcal{E}(\bmath{l}_0)&=\mathcal{P}_{d}(\bmath{l}_0, \bmath{l})-\widehat{\mathcal{P}}_{d}(\bmath{l}_0, \bmath{l}).
\end{alignat}
After calculations, the accuracy error for the $uv$-plane approximation is \ATM{simplified} to 
\begin{alignat}{2}
 \mathcal{E}(\bmath{l}_0) = & \sum_{pqkr}^{}W_{pqkr}\Big(\mathcal{P}_{pqkr}\circ\big(\mathcal{D}_{pqkr} -  \widehat{\mathcal{D}}_{pqkr}\big)\Big),\label{uverroracc}
\end{alignat}
\ATMB{while the accuracy error for} the image-plane approximation is \ATM{simplified} to
\begin{alignat}{2}
  \mathcal{E}(\bmath{l}_0) = & \PP(\textbf{0}, \bmath{l})\circ\Big(\mathcal{D}(\bmath{l}_0)- \widehat{\mathcal{D}}(\bmath{l}_0)\Big).\label{implaneerroracc}
 \end{alignat}
These analytical results of the accuracy errors indicate that the position-dependent PSFs are better approximated  using the $uv$-plane approximation approach when compared to the image-plane approach. This is easy to see in Eq.~\ref{uverroracc}; the distortion distribution is approximated separately on each baseline and per visibility  
and so does the accuracy error before the accumulation (summation) over all the \ATMB{visibilities} is carried out.
This simply means that approximating the distortion distribution for each visibility separately before the summation is carried out is more intuitive and effective than computing the summation over all the visibilities before approximating the distortion distribution based on the resulting accumulation as in Eq.~\ref{implaneerroracc}. 
As such, the $uv$-plane approximation might result in approximating the position-dependent PSFs accurately compared to the image-plane approximation. The drawback to this is that it is slower in computation as shown in \ATM{Eq.}~\ref{computimCom} when compared to the image-plane approximation. 

\begin{figure}
  \centering
    \includegraphics[width=.8\linewidth]{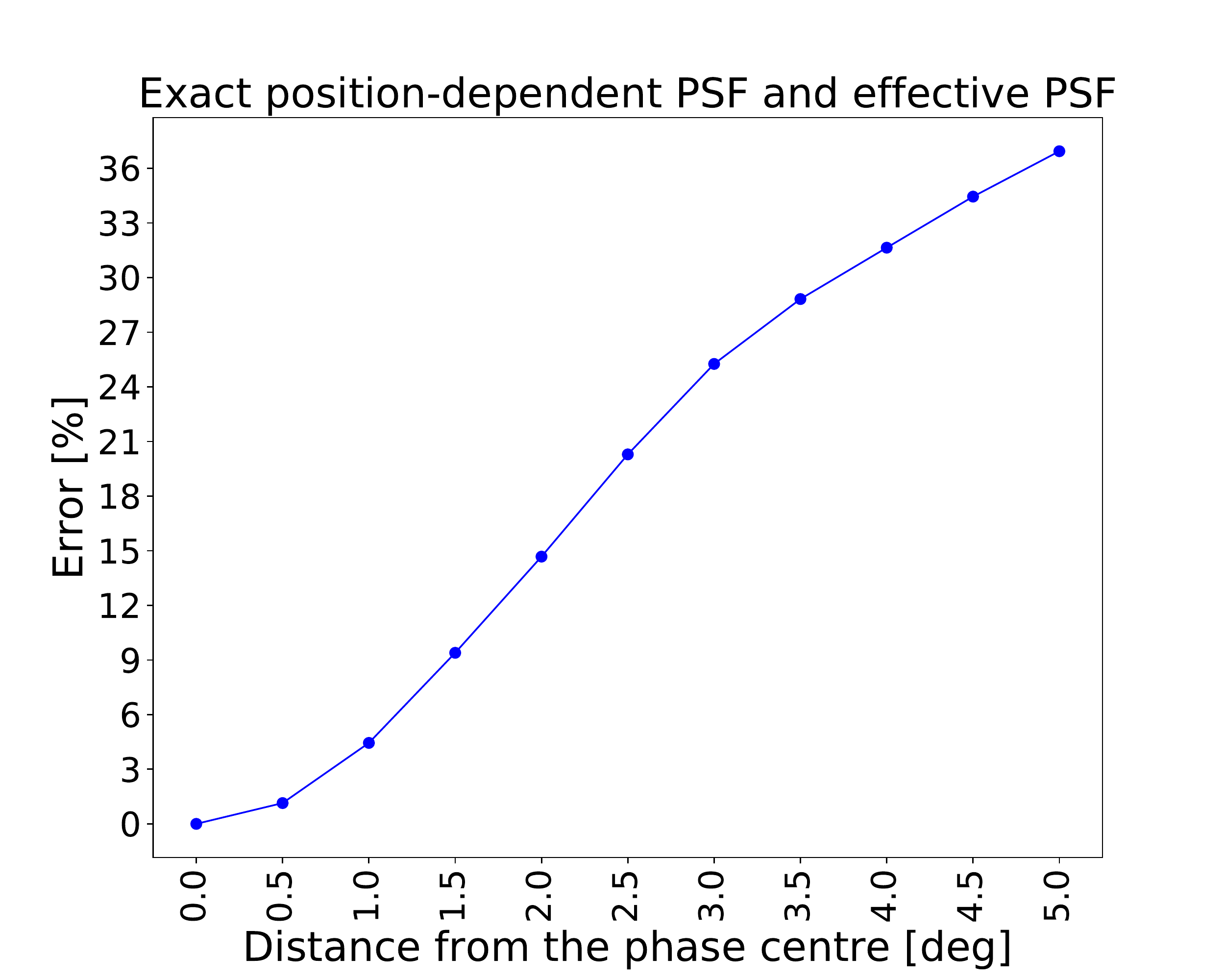}%
\caption{Maximum level of accuracy error between the exact position-dependent PSF and the effective PSF (used as its approximation) as defined in Eq.~\ref{error} (converted to a percentage) as a function of distance from the phase centre.}
\label{accuracyErrorEff}
\end{figure}

\ATM{The  accuracy error is measured by comparing the approximated position-dependent PSFs to  the exact position-dependent PSFs \ATM{computed} with brute-force.} The position-dependent PSFs are normalised and converted into decibel (dB) so one can see the differences in details between the exact and the approximated  PSFs. The  conversion in decibel: 
\begin{alignat}{2}
 &10\log_{10}|\mathcal{P}_{d}(\bmath{l}_0, \bmath{l})|~~\text{and}~~10\log_{10}|\widehat{\mathcal{P}}_{d}(\bmath{l}_0, \bmath{l})|
 \end{alignat}
 for the exact and  the approximated position-dependent PSFs respectively. \ATM{And the accuracy error in decibel is measured as:}  
 \begin{alignat}{2}
 \mathrm{Error}=&10\log_{10}|\mathcal{P}_{d}(\bmath{l}_0, \bmath{l})-\widehat{\mathcal{P}}_{d}(\bmath{l}_0, \bmath{l})|.\label{error}
\end{alignat}
\ATMB{Figure~\ref{accuracyErrorEff} shows the maximum level of the accuracy error between the exact position-dependent PSFs and the effective PSF as a function of distance from the phase centre. Since current deconvolution algorithms use the effective PSF as an \ATML{approximation} of all the exact position-dependent PSFs, it is clear from Figure~\ref{accuracyErrorEff} that the accuracy error of such approximation is very high and therefore cannot support wide-field imaging without introducing imaging artifacts. For example, a deconvolution algorithm must be able to deconvolve \ATML{without introducing} artifacts in an image with an angular distance of 0.65 deg (edge of the field of view at the FWHM of the primary beam) from the phase centre of the MeerKAT telescope at 1.4 GHz. The result in Figure~\ref{accuracyErrorEff} shows that at 0.65 deg the accuracy error is  $\sim 2.4 \%$ which is not negligible when compared to the expected accuracy error of $0 \%$ if the exact position-dependent PSF is computed.
Let us assume that one can tolerate the $2.4\%$ error at 0.65 deg, but what if  we are at the regime of wide-field imaging where for example we want to deconvolve up to $2.5$ deg (or above) from the phase centre? This will result in an accuracy error of $ \sim 20.2 \% $ which will introduce a significant amount of imaging artifacts.}

\begin{figure*}
  \includegraphics[width=1.1\textwidth]{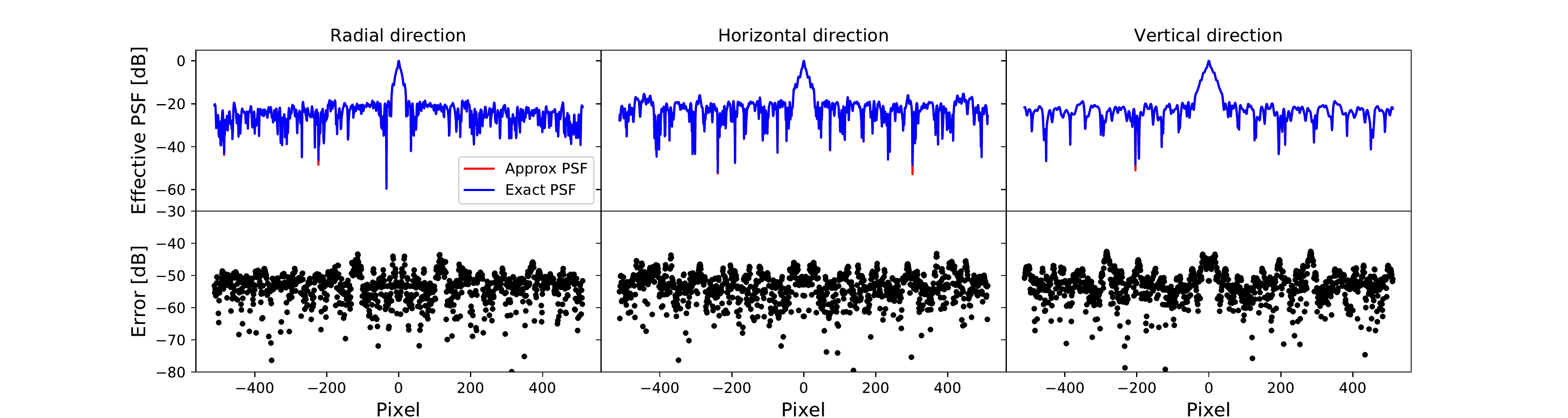}
\caption{Top panel: the effective PSF (cross-sections  in decibel)  of the MeerKAT telescope at 1.4 GHz imaged by brute-force and by approximation after averaging the visibilities from 1 s and 50 kHz to 20 s and 2 MHz; the dataset was sampled during a total time of 2 hrs and 6 MHz bandwidth. 
 Bottom panels: the accuracy error result of the subtraction between the exact and the approximated effective PSF. Only the result of the $uv$-plane approximation \ATM{is shown} in this figure because it is indeed the same result with the image-plane approximation for the effective PSF.}\label{fig:resultspsf-effectivePSFS}
\end{figure*}

\begin{figure*}
\centering
  \includegraphics[width=1.1\textwidth]{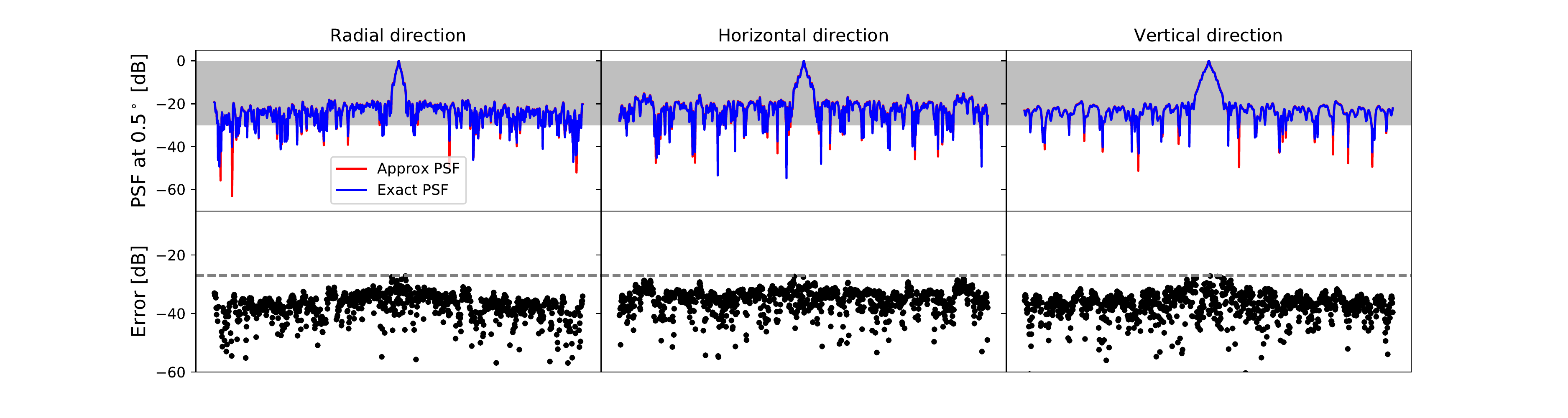}\\
  \vspace{-.5cm}
  \includegraphics[width=1.1\textwidth]{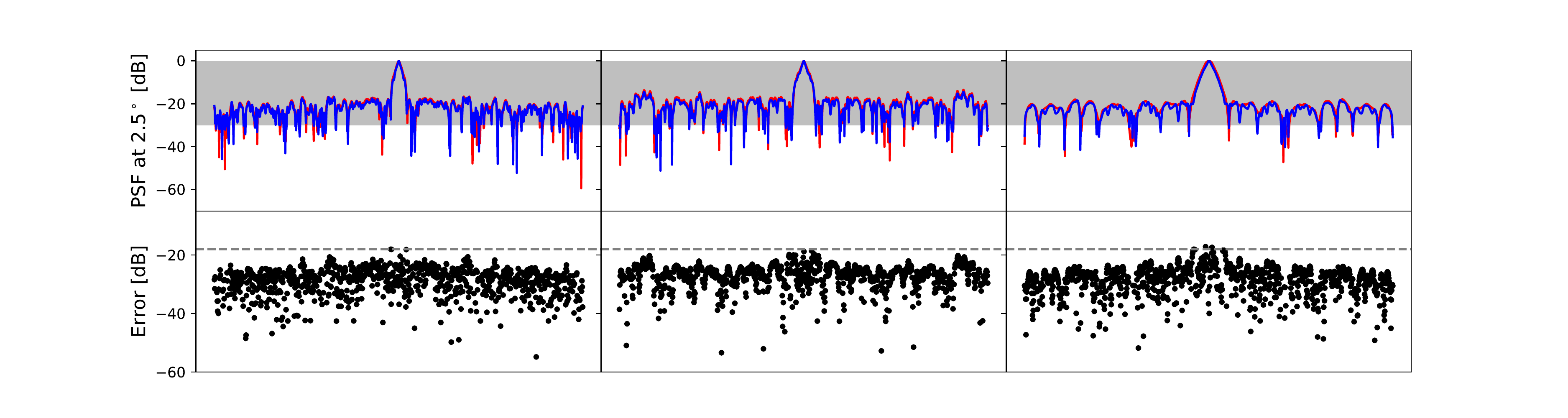}\\
  \vspace{-0.5cm}
  \includegraphics[width=1.1\textwidth]{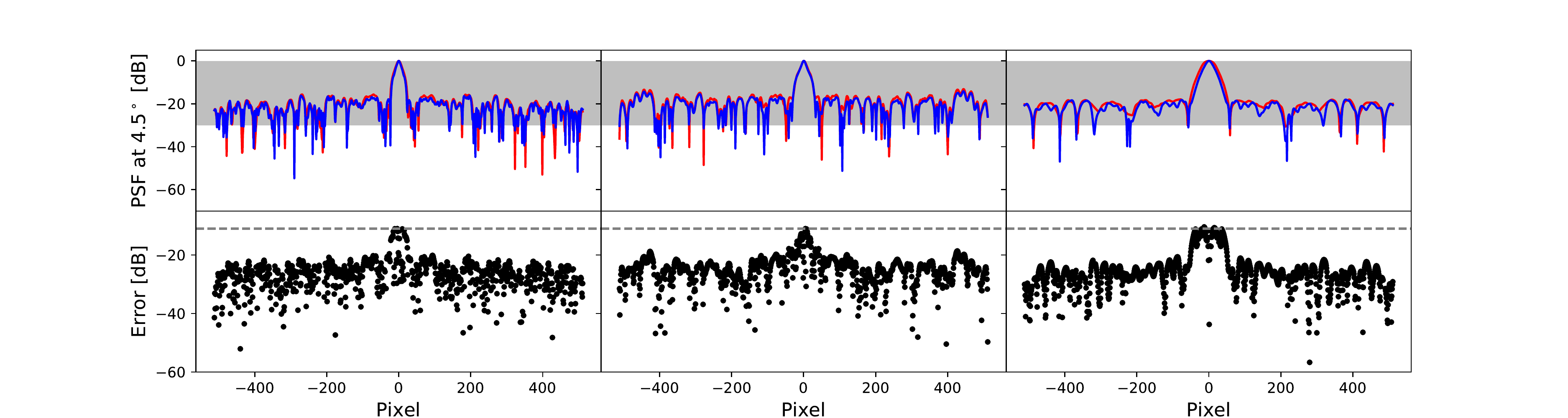}
\caption{Exact and $uv$-plane approximation of three position-dependent PSFs at $0.5$ deg (top), $2.5$ deg (middle) and $4.5$ deg (bottom) of the MeerKAT telescope at 1.4 GHz imaged after averaging the visibilities from 1 s and 50 kHz to 20 s and 2 MHz; the dataset was sampled during a total time of 2 hrs and 6 MHz bandwidth. 
The accuracy error is the result of the subtraction between the exact and the approximated position-dependent PSFs.}\label{fig:resultspsf-uv-plane}
\end{figure*}

\begin{figure*}
\centering
  \includegraphics[width=1.1\textwidth]{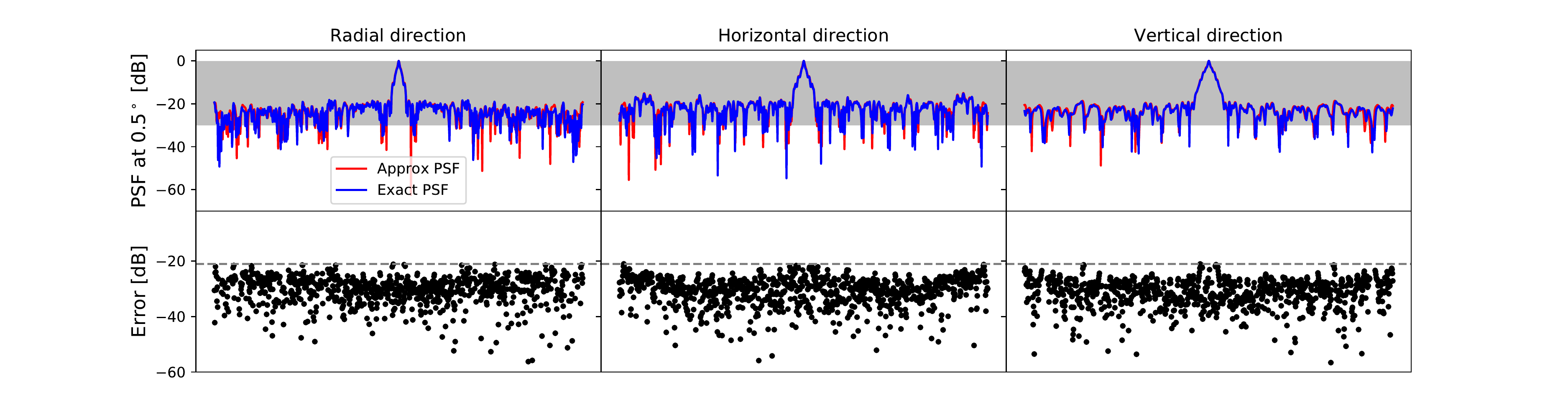}\\
  \vspace{-.5cm}
  \includegraphics[width=1.1\textwidth]{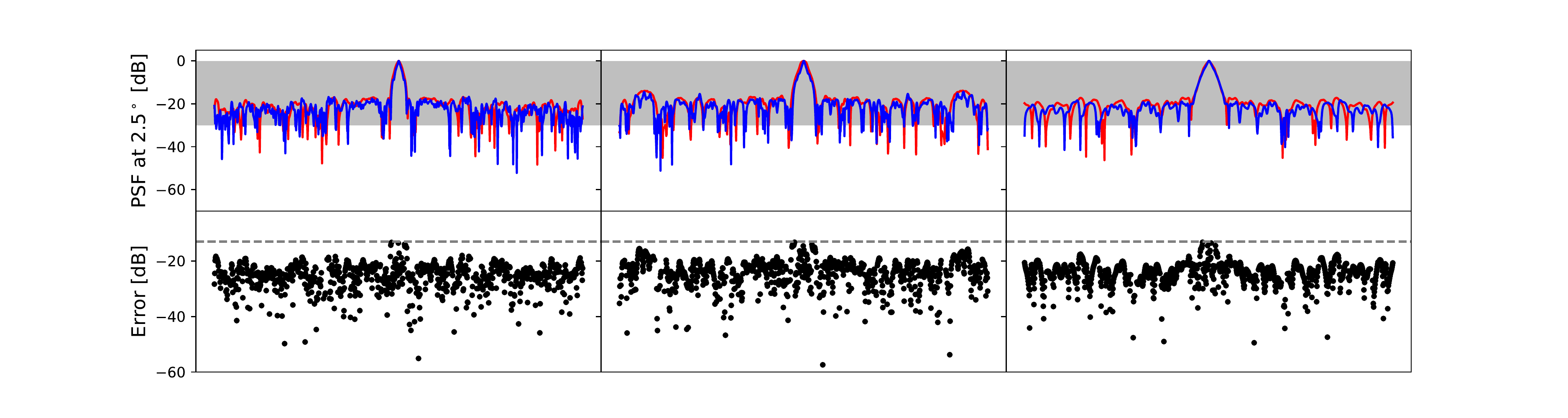}\\
  \vspace{-.5cm}
  \includegraphics[width=1.1\textwidth]{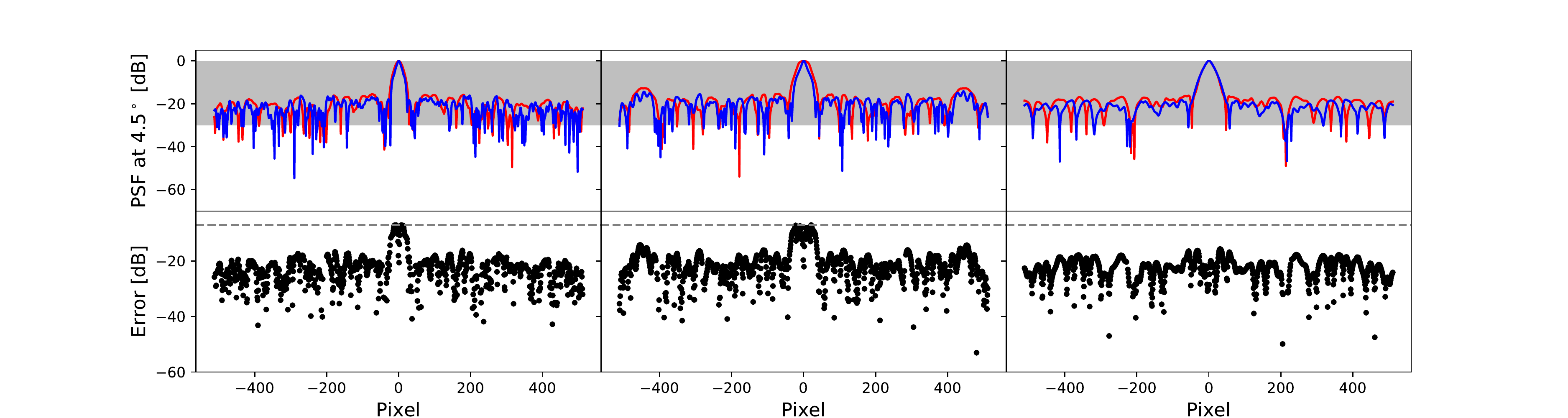}
\caption{ Exact and image-plane approximation of three position-dependent PSFs at $0.5$ deg (top), $2.5$ deg (middle) and $4.5$ deg (bottom) of the MeerKAT telescope  at 1.4 GHz imaged after averaging the visibilities from 1 s and 50 kHz to 20 s and 2 MHz; the dataset was sampled during a total time of 2 hrs and 6 MHz bandwidth. 
The accuracy error is the result of the subtraction between the exact and the approximated position-dependent PSFs.}\label{fig:resultspsf-image-plane}
\end{figure*}

Note that both proposed approximation  methods
for the position-dependent PSFs accurately approximate the effective PSF with \ATM{less than $0.01 \%$ level of accuracy error as depicted in Figure~\ref{fig:resultspsf-effectivePSFS}: the accuracy error is  below $-40$ dB}. The reason behind this is simple to understand, the visibilities for the effective PSFs have zero phase and therefore no phase decoherence after averaging and/or approximation, i.e. at 
$\bmath{l}_0=\bmath{0}$, for the $uv$-plane approximation
\begin{alignat}{2}
\forall_{pqkr}, ~&\mathcal{D}_{pqkr}(\bmath{0}) \equiv  \widehat{\mathcal{D}}_{pqkr}(\bmath{0})\sim 1
\end{alignat}
and for the image-plane approximation we have:
\begin{alignat}{2}
&\mathcal{D}(\bmath{0})\equiv \widehat{\mathcal{D}}(\bmath{0})\sim 1.
\end{alignat}
Therefore, for both approximation approaches the accuracy error becomes:
\begin{alignat}{2}
 \mathcal{E}(\bmath{l}_0)\sim 0~~\text{due to}~~\mathcal{P}_{d}(\bmath{0}, \bmath{l})\equiv \widehat{\mathcal{P}}_{d}(\bmath{0}, \bmath{l}).
\end{alignat}
The PSF is not realistic at $-30$ dB (or anything below) as one can not actually measure anything below $-30$ dB in an observation, except maybe with a calibrator source. For this reason we cutoff the normalised PSFs in decibel 
 from $-30$ dB to $0$ dB (\ATM{grey bands in Figures~\ref{fig:resultspsf-uv-plane} and \ref{fig:resultspsf-image-plane}}). 
 Figures~\ref{fig:resultspsf-uv-plane} and \ref{fig:resultspsf-image-plane} show an interesting result of the approximation, which is the differences in the main lobe and the peaks of the sidelobes between the exact and the approximated PSFs. 
 \ATM{On each of these figures, the horizontal dotted grey lines show the  \ATMB{maximum} level of accuracy error. For the $uv$-plane approximation the \ATMB{maximum} level of accuracy error is $- 27$ dB (i.e. $<0.1\%$), $- 18$ dB (i.e. $<1.5\%$) and $-11$ dB (i.e. $<7.9\%$) for source at $0.5$ deg, $2.5$ deg and $4.5$ deg respectively. And for the image-plane approximation the \ATMB{maximum} level of  accuracy error is $- 21$ dB (i.e. $<0.79\%$), $-13$ dB (i.e. $<5.1\%$) and $-7$ dB (i.e. $<19.9\%$) for a source at $0.5$ deg, $2.5$ deg and $4.5$ deg respectively.  We argue that these levels of approximation accuracy error are acceptable: \ATMB{as discussed above, assume that an accuracy error of $2.4 \%$ is acceptable. Therefore, a field with radius 0.65 deg can be deconvolved using the effective PSF as an \ATML{approximation} of the exact position-dependent PSFs because Figure~\ref{accuracyErrorEff} shows that the accuracy error is $2.4 \%$ at 0.65 deg from the phase centre. Using the $uv$-plane approximation method, for example, we can deconvolve a field with a radius of $2.5$ deg while the accuracy error is $1.5 \% < 2.4 \%$. A similar interpretation could be made with the image-plane approximation method. A field with a $2.5$ deg radius centered at the phase centre is a wide-field for the MeerKAT telescope at $1.4$ GHz.} \ATM{For a low-frequency telescope such as the LOFAR telescope, the degree of decorrelation is less when compared to a high-frequency telescope such as the MeerKAT telescope. For such a low-frequency telescope, the approximation accuracy error will be even smaller compared to the $1.5\%$  accuracy error that the MeerKAT telescope generates at $2.5$ deg centered at the phase centre. This is because the accuracy error increases with an increasing degree of decorrelation as mentioned above.   \ATMB{In this case, it is possible to deconvolve a very wide-field with low-frequency instruments while using the described methods to approximate the position-dependent PSFs: this remains an open discussion to be investigated in future works.}}}

 Note that both simulations are in agreement with the analytical interpretations: the approximation accuracy decreases (error increases) as a function of distance from the phase centre, and the $uv$-plane approximation accuracy is higher compared to the image-plane approximation accuracy.
\ATMB{The decrease in approximation accuracy} when the source is far from the phase centre is easy to understand: \ATM{there is an increasing distortion bias} around sources far from the phase centre. The approximation methods would likely be sensitive to \ATM{increased distortion bias}; e.g., the effective PSF \ATM{has zero distortion bias}, therefore, the approximation of this effective PSF results \ATM{with $\sim 0$ accuracy} error.
\section{Conclusion and future work}
\label{conclusion}
Each source in the image has its own local PSF (which we refer to as the position-dependent PSF) which are all different in amplitude and \ATMB{size}. CLEAN based algorithms use the effective PSF  to deconvolve all sources in the image. 
As a result, using the effective PSF as the position-dependent PSFs \ATML{creates} smearing artifacts that manifest differently around each source. The main reason the CLEAN based algorithms use the effective PSF as the position-dependent PSFs is that the computational requirements to compute all these position-dependent PSFs scale very poorly
as  described in Section~\ref{computationcostPSF}.
     
In this paper, we have proposed two analytical frameworks \ATM{based on an approximation} that generate these position-dependent PSFs. The first method:  the $uv$-plane approximation starts from the $uv$-plane to approximate the visibilities of the position-dependent PSFs \ATMB{from their phase gradient} before making an image. The second method: the image-plane approximation evaluates smearing coefficients at each source position and convolves with the effective PSF.
Before we started the approximation, the PSF is briefly described and \ATMB{we demonstrated that} averaging applied to visibilities for data compression purposes is the main cause that leads the PSF to be position-dependent. 
The computational cost for the two methods proposed to approximate the position-dependent PSFs is also considered. Both methods accurately approximate the position-dependent PSFs with very small accuracy error \ATM{(e.g., $<0.1\%$ and $<0.79\%$ at $0.5$ deg for the $uv$-plane and image-plane respectively)} and are less computationally demanding. Since the computational requirements using the image-plane approximation are fewer compared to the $uv$-plane approximation as shown in Section~\ref{computationcostPSF}, we recommend using the image-plane method to approximate these position-dependent PSFs during deconvolution. 

The potential of the approximation methods proposed in this work represents a step towards the development of advanced deconvolution \ATM{and/or source subtraction} techniques that use position-dependent PSFs \ATM{while using} fewer computational resources, and are capable of providing higher image fidelity.

\section*{Acknowledgements}
This work is based upon research supported by the South African Research Chairs
Initiative of the Department of Science and Technology and National Research Foundation.
The MeerKAT telescope is operated by the South African Radio Astronomy
Observatory, which is a facility of the National Research Foundation, an agency of the
Department of Science and Innovation.
MA \ATM{acknowledges} support from  Rhodes University.   \ATM{We thank Dr  Etienne Bonnassieux and Dr Chuneeta Devi Nunhokee for comments
on early drafts of this paper. We would
like to thank the anonymous referee for comments that substantially
improved the paper.}
\section*{Data availability}
The data underlying this article will be shared on reasonable request to the corresponding author.
\bibliographystyle{mn2e}
\bibliography{m_paper}

\begin{thebibliography}{}

\bibitem[\protect\citeauthoryear{Ables}{Ables}{1974}]{ables1974maximum}
Ables J.,  1974, Astronomy and Astrophysics Supplement Series, 15, 383

\bibitem[\protect\citeauthoryear{Atemkeng, Smirnov, Tasse, Foster \&
  Jonas}{Atemkeng et~al.}{2016}]{atemkeng2016}
Atemkeng M.,  Smirnov O.,  Tasse C.,  Foster G.,    Jonas J.,  2016, Monthly
  Notices of the Royal Astronomical Society, 462, 2542

\bibitem[\protect\citeauthoryear{Atemkeng, Smirnov, Tasse, Foster, Keimpema,
  Paragi \& Jonas}{Atemkeng et~al.}{2018}]{atemkeng2018baseline}
Atemkeng M.,  Smirnov O.,  Tasse C.,  Foster G.,  Keimpema A.,  Paragi Z.,
  Jonas J.,  2018, Monthly Notices of the Royal Astronomical Society, 477, 4511

\bibitem[\protect\citeauthoryear{Atemkeng}{Atemkeng}{2016}]{atemkeng2016data}
Atemkeng M.~T.,  2016, PhD thesis, RHODES UNIVERSITY

\bibitem[\protect\citeauthoryear{Bhatnagar \& Cornwell}{Bhatnagar \&
  Cornwell}{2004}]{bhatnagar2004scale}
Bhatnagar S.,  Cornwell T.,  2004, Astronomy \& Astrophysics, 426, 747

\bibitem[\protect\citeauthoryear{Bingham \& Mannila}{Bingham \&
  Mannila}{2001}]{bingham2001random}
Bingham E.,  Mannila H.,  2001, in Proceedings of the seventh ACM SIGKDD
  international conference on Knowledge discovery and data mining Random
  projection in dimensionality reduction: applications to image and text data.
pp 245--250

\bibitem[\protect\citeauthoryear{Boccardi, Krichbaum, Bach, Mertens, Ros, Alef
  \& Zensus}{Boccardi et~al.}{2016}]{boccardi2016stratified}
Boccardi B.,  Krichbaum T.,  Bach U.,  Mertens F.,  Ros E.,  Alef W.,    Zensus
  J.~A.,  2016, Astronomy \& Astrophysics, 585, A33

\bibitem[\protect\citeauthoryear{Bonnassieux, Edge, Morabito \&
  Bonafede}{Bonnassieux et~al.}{2020}]{bonnassieux2020decoherence}
Bonnassieux E.,  Edge A.,  Morabito L.,    Bonafede A.,  2020, Astronomy \&
  Astrophysics

\bibitem[\protect\citeauthoryear{Cai, Pratley \& McEwen}{Cai
  et~al.}{2019}]{cai2017online}
Cai X.,  Pratley L.,    McEwen J.~D.,  2019, Monthly Notices of the Royal
  Astronomical Society, 485, 4559

\bibitem[\protect\citeauthoryear{Carrillo, McEwen \& Wiaux}{Carrillo
  et~al.}{2014}]{carrillo2014purify}
Carrillo R.~E.,  McEwen J.~D.,    Wiaux Y.,  2014, Monthly Notices of the Royal
  Astronomical Society, 439, 3591

\bibitem[\protect\citeauthoryear{Cooley \& Tukey}{Cooley \&
  Tukey}{1965}]{cooley1965algorithm}
Cooley J.~W.,  Tukey J.~W.,  1965, Mathematics of computation, 19, 297

\bibitem[\protect\citeauthoryear{Dabbech, Ferrari, Mary, Slezak, Smirnov \&
  Kenyon}{Dabbech et~al.}{2015}]{dabbech2015moresane}
Dabbech A.,  Ferrari C.,  Mary D.,  Slezak E.,  Smirnov O.,    Kenyon J.~S.,
  2015, Astronomy \& Astrophysics, 576, A7

\bibitem[\protect\citeauthoryear{Dewdney, Hall, Schilizzi \& Lazio}{Dewdney
  et~al.}{2009}]{dewdney2009square}
Dewdney P.~E.,  Hall P.~J.,  Schilizzi R.~T.,    Lazio T. J.~L.,  2009,
  Proceedings of the IEEE, 97, 1482

\bibitem[\protect\citeauthoryear{Golub \& Reinsch}{Golub \&
  Reinsch}{1970}]{golub1970singular}
Golub G.~H.,  Reinsch C.,  1970, Numerische mathematik, 14, 403

\bibitem[\protect\citeauthoryear{H{\"o}gbom}{H{\"o}gbom}{1974}]{hogbom1974aperture}
H{\"o}gbom J.,  1974, Astronomy and Astrophysics Supplement Series, 15, 417

\bibitem[\protect\citeauthoryear{Jonas}{Jonas}{2009}]{jonas2009meerkat}
Jonas J.~L.,  2009, Proceedings of the IEEE, 97, 1522

\bibitem[\protect\citeauthoryear{Junklewitz, Bell, Selig \&
  En{\ss}lin}{Junklewitz et~al.}{2016}]{junklewitz2016resolve}
Junklewitz H.,  Bell M.,  Selig M.,    En{\ss}lin T.,  2016, Astronomy \&
  Astrophysics, 586, A76

\bibitem[\protect\citeauthoryear{Kartik, Carrillo, Thiran \& Wiaux}{Kartik
  et~al.}{2017}]{vijay2017fourier}
Kartik S.,  Carrillo R.,  Thiran J.,    Wiaux Y.,  2017, Monthly Notices of the
  Royal Astronomical Society, 468, 2382

\bibitem[\protect\citeauthoryear{Labate, Braun, Dewdney, Waterson \&
  Wagg}{Labate et~al.}{2017}]{labate2017ska1}
Labate M.,  Braun R.,  Dewdney P.,  Waterson M.,    Wagg J.,  2017, in 2017
  XXXIInd General Assembly and Scientific Symposium of the International Union
  of Radio Science (URSI GASS) Ska1-low: Design and scientific objectives.
pp~1--4

\bibitem[\protect\citeauthoryear{Meillier, Ammanouil, Ferrari \&
  Bianchi}{Meillier et~al.}{2018}]{meillier2018distribution}
Meillier C.,  Ammanouil R.,  Ferrari A.,    Bianchi P.,  2018, Signal
  Processing: Image Communication

\bibitem[\protect\citeauthoryear{Offringa}{Offringa}{2016}]{offringa2016compression}
Offringa A.,  2016, Astronomy \& Astrophysics, 595, A99

\bibitem[\protect\citeauthoryear{Offringa et~al.,}{Offringa
  et~al.}{2014}]{offringa2014wsclean}
Offringa A.,  et~al., 2014, Monthly Notices of the Royal Astronomical Society,
  444, 606

\bibitem[\protect\citeauthoryear{Smirnov}{Smirnov}{2011}]{smirnov2011revisiting}
Smirnov O.~M.,  2011, Astronomy \& Astrophysics, 527, A107

\bibitem[\protect\citeauthoryear{Smith, Young \& Davidson}{Smith
  et~al.}{2017}]{smith2017reconsidering}
Smith D.,  Young A.,    Davidson D.,  2017, Astronomy \& Astrophysics, 603, A40

\bibitem[\protect\citeauthoryear{Tasse, Hugo, Mirmont, Smirnov, Atemkeng,
  Bester, Hardcastle, Lakhoo, Perkins \& Shimwell}{Tasse
  et~al.}{2018}]{tasse2018faceting}
Tasse C.,  Hugo B.,  Mirmont M.,  Smirnov O.,  Atemkeng M.,  Bester L.,
  Hardcastle M.,  Lakhoo R.,  Perkins S.,    Shimwell T.,  2018, Astronomy \&
  Astrophysics, 611, A87

\bibitem[\protect\citeauthoryear{Thompson}{Thompson}{1999}]{thompson1999fundamentals}
Thompson A.~R.,  1999, in {Synthesis Imaging in Radio Astronomy II} Vol.~180,
  {Fundamentals of radio interferometry}.
p.~11

\bibitem[\protect\citeauthoryear{{Thompson}, {Moran} \& {Swenson,
  Jr.}}{{Thompson} et~al.}{2001}]{thompson2001fundamentals}
{Thompson} A.~R.,  {Moran} J.~M.,    {Swenson, Jr.} G.~W.,  2001,
  {Interferometry and Synthesis in Radio Astronomy}, 2 edn.
Wiley, {New York}

\bibitem[\protect\citeauthoryear{van Haarlem et~al.,}{van Haarlem
  et~al.}{2013}]{van2013lofar}
van Haarlem M.~P.,  et~al., 2013, Astronomy \& astrophysics, 556, A2

\bibitem[\protect\citeauthoryear{Wijnholds, Willis \& Salvini}{Wijnholds
  et~al.}{2018}]{wijnholds2018baseline}
Wijnholds S.,  Willis A.,    Salvini S.,  2018, Monthly Notices of the Royal
  Astronomical Society, 476, 2029

\bibitem[\protect\citeauthoryear{Ye, Gull, Tan \& Nikolic}{Ye
  et~al.}{2020}]{ye2020optimal}
Ye H.,  Gull S.~F.,  Tan S.~M.,    Nikolic B.,  2020, Monthly Notices of the
  Royal Astronomical Society, 491, 1146

\end{thebibliography}

\bsp
\end{document}